\documentclass[aps,pra,twocolumn,amssymb,amsfonts,floatfix,superscriptaddress,longbibliography]{revtex4-1}

\pdfoutput=1

\usepackage[applemac]{inputenc}
\usepackage[english]{babel}
\usepackage[pdftex]{graphicx}
\usepackage{color}
\usepackage{multirow}
\usepackage{dcolumn}
\usepackage{bm}
\usepackage{mathbbol}
\usepackage{xcolor}
\usepackage{amsmath,amsthm}
\usepackage{times}
\usepackage[normalem]{ulem}

\begin{document}
\title{Dynamical signatures of quantum chaos and relaxation timescales in a spin-boson system}
\author{S. Lerma-Hern\'andez}
\affiliation{Facultad de F\'isica, Universidad Veracruzana, Circuito Aguirre Beltr\'an s/n, Xalapa, Veracruz 91000, Mexico}
 \author{D. Villase\~nor}
 \affiliation{Instituto de Ciencias Nucleares, Universidad Nacional Aut\'onoma de M\'exico, Apdo. Postal 70-543, C.P. 04510  Cd. Mx., Mexico}
\author{M. A. Bastarrachea-Magnani}
\affiliation{Department of Physics and Astronomy, Aarhus University, Ny Munkegade, DK-8000 Aarhus C, Denmark.}
\author{E. J. Torres-Herrera} 
\affiliation{Instituto de F\'isica, Benem\'erita Universidad Aut\'onoma de Puebla, Apdo. Postal J-48, Puebla, Puebla 72570, Mexico}
 \author{L. F. Santos} 
\affiliation{Department of Physics, Yeshiva University, New York, New York 10016, USA}
 \author{J. G. Hirsch} 
\affiliation{Instituto de Ciencias Nucleares, Universidad Nacional Aut\'onoma de M\'exico, Apdo. Postal 70-543, C.P. 04510  Cd. Mx., Mexico}
%%%%%%%%%%%%%%%%%%%%%%%%%%%%%%%%%%%%%%%% 

\begin{abstract}
Quantum systems whose classical counterparts are chaotic typically  have highly correlated eigenvalues and level statistics that coincide with those from ensembles of full random matrices. A dynamical manifestation of these correlations comes in the form of the so-called correlation hole, which is a dip below the saturation point of the survival probability's time evolution. In this work, we study the correlation hole in the spin-boson (Dicke) model, which presents a chaotic regime and can be realized in experiments with ultracold atoms and ion traps. We derive an analytical expression that describes the entire evolution of the survival probability and allows us to determine the timescales of its relaxation to equilibrium. This expression shows remarkable agreement with our numerical results. While the initial decay and the time to reach the minimum of the correlation hole depend on the initial state, the dynamics beyond the hole up to equilibration is universal. We find that the relaxation time of the survival probability for the Dicke model increases linearly with system size. 
\end{abstract}

\maketitle

%%%%%%%%%%%%%%%%%%%%%%%%%%%%%%%%%%%%%%%%
%%%%%%%%%%%%%%%%%%%%%%%%%%%%%%%%%%%%%%%%

\section{Introduction}
\label{sec01}

The subject of equilibration and thermalization of isolated quantum systems in the chaotic regime has seen a great deal of advance in the last years~\cite{Reimann2008,Short2011,Short2012,Zangara2013,HeSantos2013,Gogolin2016,Borgonovi2016,Dalessio2016,DymarskyARXIV,Reimann2018a,*Reimann2018b}. Equilibration is reached when, after a transient time, the observable under investigation shows only small fluctuations around an asymptotic value, and these fluctuations decrease with system size. Thermalization implies that this infinite-time average is very close to the predictions from statistical mechanics, and the difference between the two also decreases with system size. In this picture, an important open question is how long it takes for isolated quantum systems to reach equilibrium. Despite the increasing number of recent works addressing this issue~\cite{Monnai2013,Goldstein2013,Malabarba2014,Goldstein2015,Gogolin2016,Reimann2016,Pintos2017,Oliveira2018,DymarskyARXIVThouless,Chan2018,Bertini2018,SchiulazARXIV,Borgonovi2019}, there is no agreement regarding how the relaxation timescale should depend on system size, range of interactions, observables, and initial states.

The last steps of the evolution toward equilibrium, after the dynamics resolves the discreteness of the spectrum, are determined by the properties of the eigenvalues~\cite{SchiulazARXIV}. The largest possible timescale is the Heisenberg time~\cite{BerryProceed1981}, which is proportional to the inverse of the mean level spacing of the region of the spectrum probed by the initial state. Before reaching this timescale, effects of the correlations between the eigenvalues may be observed. In the case of the survival probability~\cite{footSurv}, which is the probability of finding the initial state at time $t$, these correlations cause a decay below the saturation value of the dynamics, known as correlation hole~\cite{footNote}. This phenomenon was first studied in the context of molecules, where the interest was not exactly in dynamics, but in alternative ways to detect level repulsion in systems without good line resolution~\cite{Leviandier1986,Guhr1990,Wilkie1991}. 

The correlation hole has been studied in full random matrices~\cite{Alhassid1992}, in many-body systems with~\cite{Torres2017,Torres2017Philo,Torres2018,Torres2019,SchiulazARXIV} and without disorder~\cite{Torres2017Philo}, in the Sachdev-Ye-Kitaev model~\cite{Cotler2017GUE,Gharibyan2018,Nosaka2018}, which is a two-body random ensemble~\cite{Brody1981}, and in the finite one-dimensional Anderson model~\cite{TorresARXIV}. The hole is not exclusive to the survival probability, but emerges also in experimental local observables~\cite{Torres2018,Torres2019}. For the correlation hole to be visible, one needs to perform large averages over initial states and, in the case of Hamiltonian matrices with random elements, over ensembles of Hamiltonian realizations. In Ref.~\cite{SchiulazARXIV}, it was shown that in realistic chaotic many-body quantum systems with local short-range interactions and perturbed far from equilibrium, the time to reach the minimum of the correlation hole increases exponentially with system size. This timescale, which is still shorter than the Heisenberg time, was referred to as Thouless time due to its relationship with the Thouless energy computed from random matrix theory. As explained in~\cite{SchiulazARXIV}, the Thouless time in  interacting systems is the time that it takes for an initial state to spread over the entire Hilbert space accessible to its energy. Beyond this point, the dynamics becomes universal all the way to equilibrium. 

In the present work, we use the survival probability to study the correlation hole in the Dicke model. This is a paradigmatic spin-boson model with two degrees of freedom. It has a classical counterpart and exhibits chaos for several values of its parameters, mostly for high excitation energies and in the superradiant phase. The model was first introduced to explain superradiance~\cite{Dicke1954,Garraway2011,Baumann2010,Baumann2011} and has since then been used in different contexts, from quantum chaos~\cite{Lewenkopf1991,Emary2003PRL,*Emary2003,Bastarrachea2014b,*Bastarrachea2015,*Bastarrachea2016PRE,Chavez2016} and quantum batteries~\cite{AndolinaARXIV} to excited-state quantum phase transitions~\cite{Fernandez2011b,Brandes2013,Bastarrachea2014a,Larson2017} and quench dynamics~\cite{Fernandez2011,Altland2012PRL,Lerma2018,Kloc2018}. Recently, the model was employed in a study of the out-of-time ordered correlator (OTOC), where it was shown that, in the chaotic regime, the OTOC increases exponentially in time with a rate comparable to the classical Lyapunov exponent~\cite{Chavez2019}. In addition to ultracold atoms in optical cavities~\cite{Baden2014,Klinder2015}, the Dicke model can now be realized also with ion traps 
 \cite{Cohn2018}. The latter is one of the main platforms to study long-time coherence evolution~\cite{Blatt2012}, which makes the analysis of the timescales involved in the relaxation process of the Dicke model a timely subject.

We obtain an analytical expression that describes the entire evolution of the survival probability for the Dicke model in the chaotic regime. The expression describes very accurately our numerical results, and with it, we can derive analytically the timescales involved in the relaxation process. We find that the relaxation time increases linearly with system size, while the Thouless time depends non trivially on the initial state.

The article is organized as follows. In Sec.~\ref{sec02}, we describe the Dicke model and the properties associated with its eigenvalues.  In Sec.~\ref{sec03}, we discuss the initial states considered and present the analytical expression for the survival probability. This expression is compared with numerical results in Sec.~\ref{sec04}. The analytical expressions for the Thouless and relaxation times are given and discussed in Sec.~\ref{ssec02d}. We present our conclusions in Sec.~\ref{sec06}.

%%%%%%%%%%%%%% SEC.2  DICKE MODEL %%%%%%%%%%%%%%%%

\section{Dicke Model}  
\label{sec02}

The Dicke model~\cite{Dicke1954}  describes the interaction between a set of $\mathcal{N}$ two-level atoms with energy splitting $\omega_{0}$ and a  single mode of the electromagnetic field with radiation frequency $\omega$. By setting $\hbar=1$, the time-reversal symmetric Hamiltonian of the model is written as
\begin{equation}
\label{eqn01}
\hat{H}_{D}=\omega\hat{a}^{\dagger}\hat{a}+\omega_{0}\hat{J}_{z}+\frac{2\gamma}{\sqrt{\mathcal{N}}}\hat{J}_{x}(\hat{a}^{\dagger}+\hat{a}).
\end{equation}
The first term of the equation above accounts for the energy of the field, where $\hat{a}^{\dagger}$ ($\hat{a}$) is the bosonic creation (annihilation) operator. The second term corresponds to the energy of the atoms, where $\hat{J}_{x,y,z}=\frac{1}{2}\sum_{k=1}^{\mathcal{N}}\sigma_{x,y,z}^{k}$ are the atomic pseudo-spin operators and $\sigma_{x,y,z}$ are the Pauli matrices. The third term describes the atom-field interaction with coupling parameter $\gamma$. The eigenvalues $j(j+1)$ of the operator $\hat{\textbf{J}}^{2}=\hat{J}_{x}^{2}+\hat{J}_{y}^{2}+\hat{J}_{z}^{2}$  determine different invariant subspaces. Its maximum value, given by $j=\mathcal{N}/2$, defines the symmetric non-degenerate atomic subspace that includes the ground-state. The Hamiltonian $\hat{H}_{D}$ commutes with the parity operator $\hat{\Pi}=e^{i\pi\hat{\Lambda}}$, 
where $\hat{\Lambda}=\hat{a}^{\dagger}\hat{a}+\hat{J}_{z}+j\hat{1}$. The operator $\hat{\Lambda}$ represents the total number of excitations with eigenvalues  $\lambda=n+m+j$, where $n$ is the number of photons, $m+j$ is the number of excited atoms, and  $m$ is the eigenvalue of $\hat{J}_z$. In all calculations presented below, we consider the positive parity spectrum of the model.

When the coupling parameter reaches a critical value $\gamma_{c}=\sqrt{\omega\omega_{0}}/2$, a second-order quantum phase transition takes place~\cite{Hepp1973a,Hepp1973b}. The system goes from  a normal phase ($\bar{\gamma}<1$ with  $\bar{\gamma}=\gamma/\gamma_{c}$), where the ground state has no photons and all the atoms are in their lowest level, to a superradiant phase ($\bar{\gamma}>1$), where the ground state has non-zero expectation values for the number of photons and number of excited atoms.

\subsection{Level Statistics and Density of States}
The classical limit of the Dicke model can be obtained by using Bloch and Glauber coherent states~\cite{Bastarrachea2014a,Bastarrachea2014b,Bastarrachea2015,Chavez2016}, which allows for the identification of the parameters and energy range that lead to chaos.  A main signature of classical chaos in the quantum regime is energy-level repulsion~\cite{Casati1980,Bohigas1984}. 

\begin{figure}[htb]
\centering{
\vskip 0.4 cm 
\includegraphics[width=.45\textwidth]{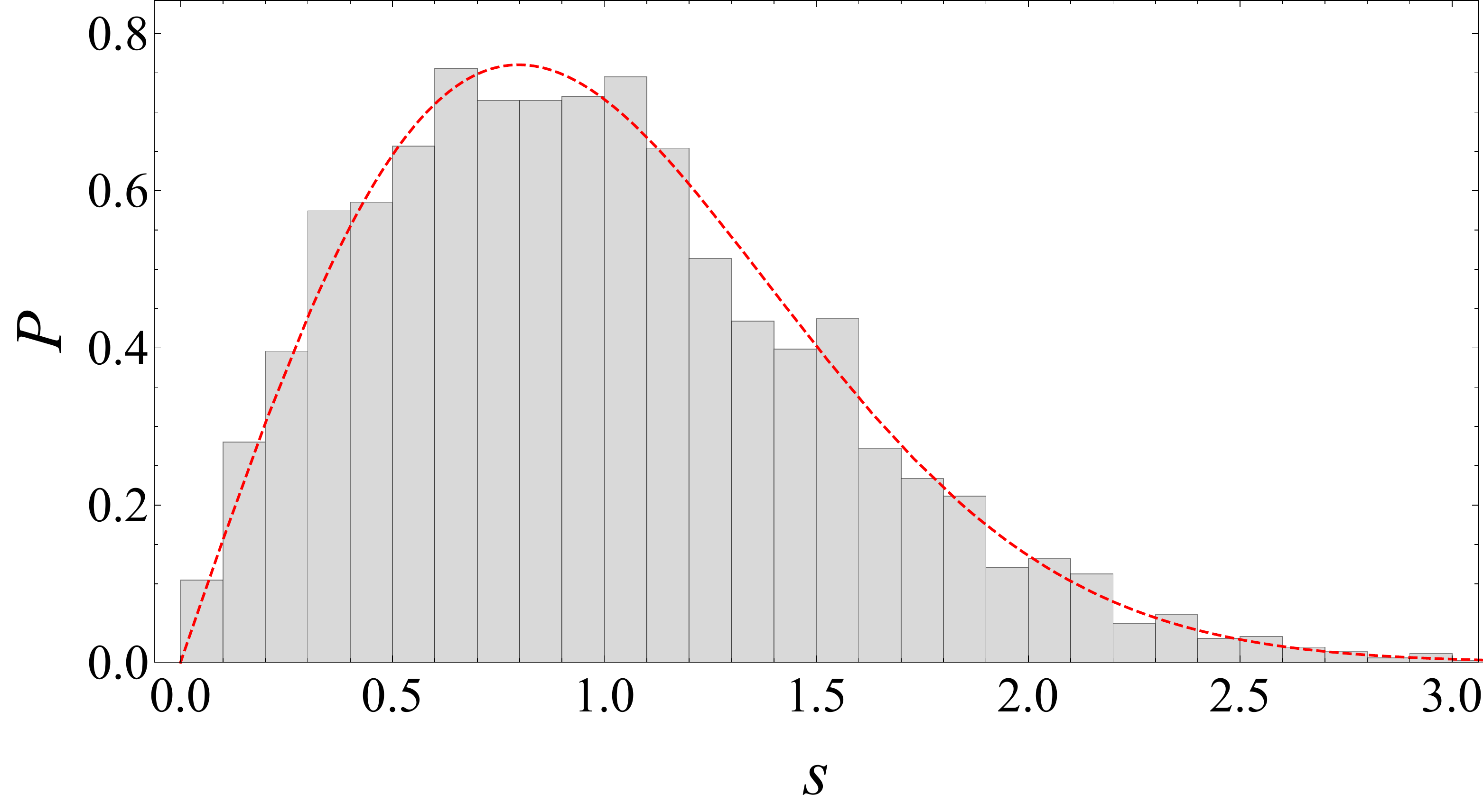}  \vskip - 4.3 cm \hspace{-5.5 cm} (a)  \\
\vskip 4.3 cm 
\hspace{-5.5 cm}  (b) \\
\vskip -0.7 cm 
\includegraphics[width=8cm]{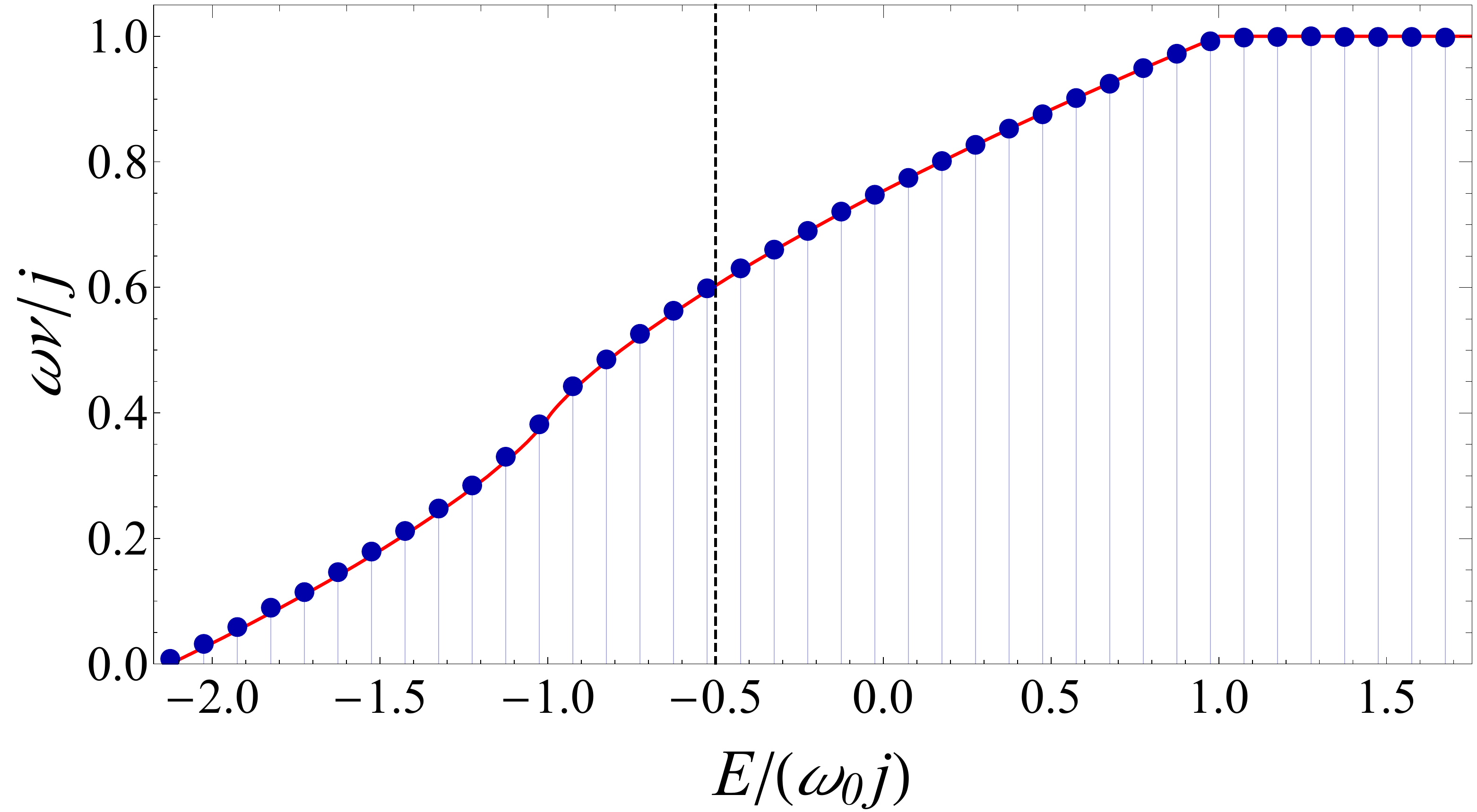} }
\caption{ (Color online) In  (a): The level spacing distribution for the unfolded spectrum (shaded area) in the energy region $E/(\omega_0 j) \in [-0.8, -0.2]$ agrees with the Wigner surmise (dashed line). 
 In (b): Density of states (DoS) evaluated numerically (blue circles) with bin size $\Delta E=0.1 \omega_0 j$ and analytical expression (\ref{eqn02}) indicated with the red solid curve. Parameters: $\omega=\omega_{0}$, $\bar{\gamma}=2$, $j=100$, and positive parity. The vertical line in (b) indicates the energy $E_c=-0.5\omega_0 j$ chosen for our study.  A truncated Hilbert space was employed   using the basis of Refs.~\cite{Bastarrachea2014PSa,Bastarrachea2014PSb},  ensuring 24\,453 converged eigenenergies, which range from the ground state energy $E_{gs}=-2.125 \omega_0 j$ until $E_{T}=1.755\omega_0 j$.
}
\label{figS2}
\end{figure}

As our case study, we choose $\omega=\omega_0$ and $j=100$, and we select a coupling parameter in the superradiant phase, $\bar{\gamma}=2$.  For these values, chaos is found at excitation energies above $E\approx -1.6 \omega_0 j$ (see Ref.~\cite{Chavez2016}). We choose an energy well above this threshold, $E_c=-0.5 \omega_0 j$, for which the whole energy shell is covered by chaotic trajectories (see Ref.~\cite{Chavez2016}). In  Fig.~\ref{figS2}~(a), we show the level spacing distribution, denoted by $P(s)$, where $s$ is the spacing between nearest-neighboring unfolded energy levels from an energy interval around $E_c$. In quantum systems whose classical counterparts are chaotic, the levels are prohibited from crossing and $P(s)$ coincides with the Wigner surmise~\cite{Guhr1998}, as indeed confirmed in Fig.~\ref{figS2}~(a). 

With the classical Hamiltonian, it is possible to estimate the energy averaged density of states (DoS), which is given by the expression~\cite{Bastarrachea2014a},
\begin{equation}
\nu(E)=\frac{j}{\omega}\left\{\begin{array}{l}\frac{1}{\pi}\int_{y_{-}}^{y_{+}}dy\cos^{-1}\left(\sqrt{\frac{2(y-\epsilon)}{\bar{\gamma}^{2}(1-y^{2})}}\right), \epsilon_{gs}\leq\epsilon<-1  \\ 
\frac{1+\epsilon}{2}+\frac{1}{\pi}\int_{\epsilon}^{y_{+}}dy\cos^{-1}\left(\sqrt{\frac{2(y-\epsilon)}{\bar{\gamma}^{2}(1-y^{2})}}\right), |\epsilon|\leq 1 \\ 
1, \epsilon>1\end{array}\right.
\label{eqn02}
\end{equation}
where $y_{\pm}=-\bar{\gamma}^{-1}\left(\bar{\gamma}^{-1}\mp\sqrt{2(\epsilon-\epsilon_{gs})}\right)$ and $\epsilon=\frac{E}{\omega_{0}j}$ is the normalized energy. The ground state energy is $\epsilon_{gs}=-1$  for the normal phase, while it is $\epsilon_{gs}=-\dfrac{1}{2}\left(\bar{\gamma}^{2}+\bar{\gamma}^{-2}\right)$ in the superradiant phase. In Fig.~\ref{figS2}~(b), we compare the DoS obtained numerically with the expression (\ref{eqn02}). The agreement is excellent. It is evident from the figure that for $|\epsilon|\leq 1$, the DoS shows a linear dependence on energy, $\nu(E) \propto E$. Our choice of the value of $E_c $  for the studies below falls within this region. Notice also that the DoS in Eq.~(\ref{eqn02}) scales linearly with the number of atoms ($j$ appears explicitly in the beginning of the equation), a property that will be useful below to determine the dependence of the timescales of the model on the number of atoms. 

%%%%%%%%%%%% SEC.3   SURVIVAL PROBABILITY %%%%%%%%%%%%%
\section{Survival probability and Initial states}
\label{sec03}

The survival probability, $S_{P}(t)$, is a dynamical observable defined as the probability to find an arbitrary initial  quantum state $|\Psi (0) \rangle$  at a later time $t$,
\begin{equation}
\label{eqn03}
S_{P}(t)=|\langle\Psi(0)|\Psi(t)\rangle|^{2}.
\end{equation}
By writing the initial state in terms of the energy eigenbasis, $|\Psi (0) \rangle=\sum_k c_k |\phi_k\rangle$, where $\hat{H}|\phi_{k}\rangle=E_{k}|\phi_{k}\rangle$ and  $c_{k}=\langle\phi_{k}|\Psi(0)\rangle$, the survival probability is
\begin{equation}
S_{P}(t)=\left|\sum_{k}|c_{k}|^{2}e^{-i E_{k}t}\right|^{2} .
\label{Eq:SP}
\end{equation}
The short-time evolution depends on the energy distribution of the initial state~\cite{Lerma2018,Torres2014PRA}, while the long-time dynamics is determined by the properties of the spectrum~\cite{Torres2017Philo,Torres2018}. The survival probability has been studied in several different contexts, with early works focusing on deviations from exponential behaviors~\cite{Khalfin1958,Fonda1978} and the quantum speed limit~\cite{Bhattacharyya1983}.

\subsection{Initial States}

To disentangle the effects of the spectrum from those of the energy components of the initial state in the behavior of the survival probability, we consider ensembles of initial states defined in a given chaotic energy region with components randomly selected, so that
\begin{equation}
|c_k|^2= \frac{r_k f(E_k)}{\sum_q r_q f(E_q)}.
\label{eq:components}
\end{equation} 
Above, $r_k$ are positive random numbers  from an arbitrary probability distribution. For the numerical simulations presented   below, we consider an  uniform distribution in the interval $[0,1]$ with $n$-th  moments $\langle r_k^n\rangle= 1/(n+1)$. The function $f(E)=\rho(E)/\nu(E)$ is used to guarantee that the initial state has a certain selected profile $\rho(E)$, which is achieved by compensating for changes in the density of states. We consider normalized rectangular and  Gaussian profiles given respectively by,
\begin{eqnarray}
\rho^R(E)&=& \left\{\begin{array}{lr}\dfrac{1}{2\sigma_R}&  {\hbox  {  \  \ \  \  \ \ \ \ \ for $E \in [E_c-\sigma_R,E_c+\sigma_R]$}} 
\\
                              0&  {\hbox  { otherwise}} , \end{array} \right.
                              \label{Eq:rect}
\\ 
\rho^G(E)&=& \left\{\begin{array}{lr} \dfrac{e^{-(E-E_c)^2/(2 \sigma_G^2)}}{{\cal C}\sigma_G \sqrt{2\pi }}
&
 {\hbox  { for $E \in [E_{\textrm{min}},E_{\textrm{max}}]$}} 
\\
                              0&  {\hbox  { otherwise}}. \end{array} \right.
\label{Eq:Gauss}
\end{eqnarray}
The profiles are centered at the energy $E_c$, where we know that chaos dominates the dynamics. The widths of the rectangular and Gaussian profiles are, respectively, $\sigma_R$ and $\sigma_G$. The lower and upper energy bounds of the Gaussian profile are $E_{\textrm{min}}$ and $E_{\textrm{max}}$, and ${\cal C}$ is a normalization factor,
\begin{equation}
{\cal C}=\frac{1}{2}\left[\textrm{erf}\left(\frac{E_\textrm{c}-E_\textrm{min}}{\sqrt{2}\sigma_G }\right)-\textrm{erf}\left(\frac{E_\textrm{c}-E_\textrm{max}}{\sqrt{2}\sigma_G }\right)\right],
\end{equation}
with $\textrm{erf}$ being the error function.
In the context of quench dynamics, where the system is initially prepared in a coherent state, the energy distribution of the initial state is indeed Gaussian, which makes the Gaussian profile a realistic choice (for some examples, see Ref.~\cite{Lerma2018}). The bounds $E_{\textrm{min}}$ and $E_{\textrm{max}}$, especially $E_{\textrm{min}}$, are also plausible, since in quantum systems there is always at least a ground state, whose presence should affect the dynamics by partially reconstructing the initial state~\cite{MugaBook,Tavora2016,Tavora2017}.

\begin{figure*}[htb]
\begin{tabular}{ccc}
(a)& (b)&  (c) \\
\includegraphics[width=.32\textwidth]{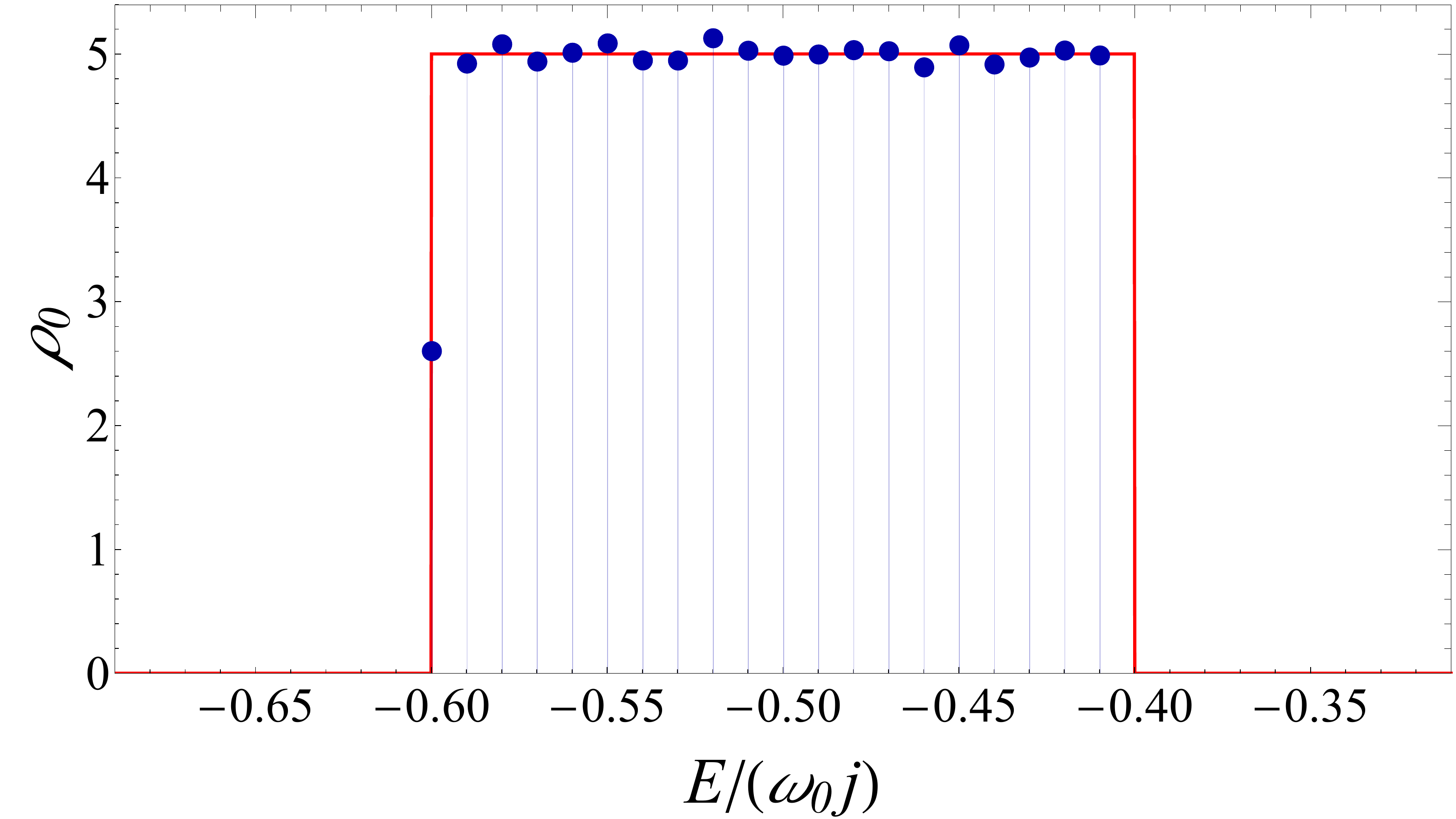}&
\includegraphics[width=.32\textwidth]{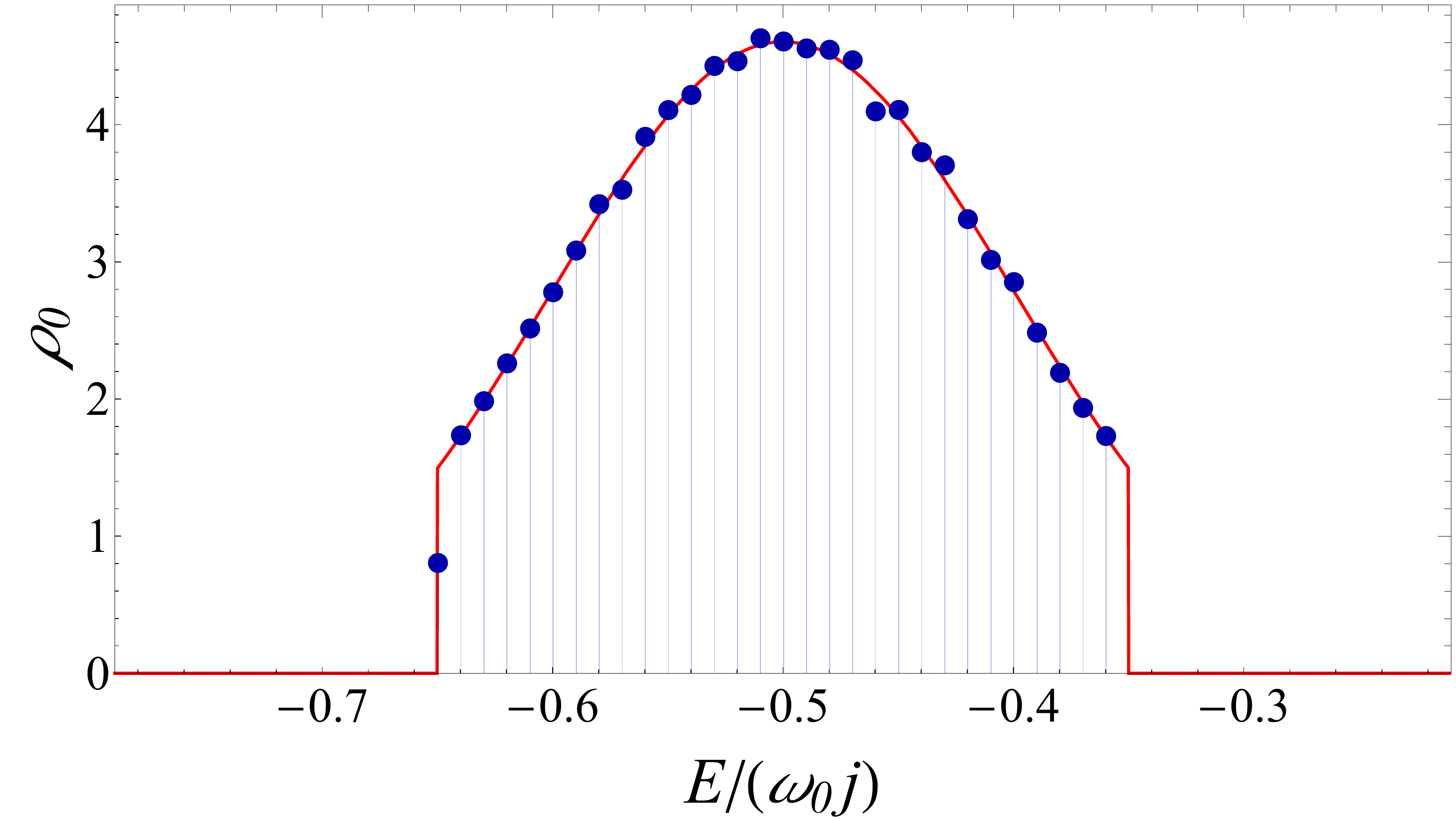}&
\includegraphics[width=.32\textwidth]{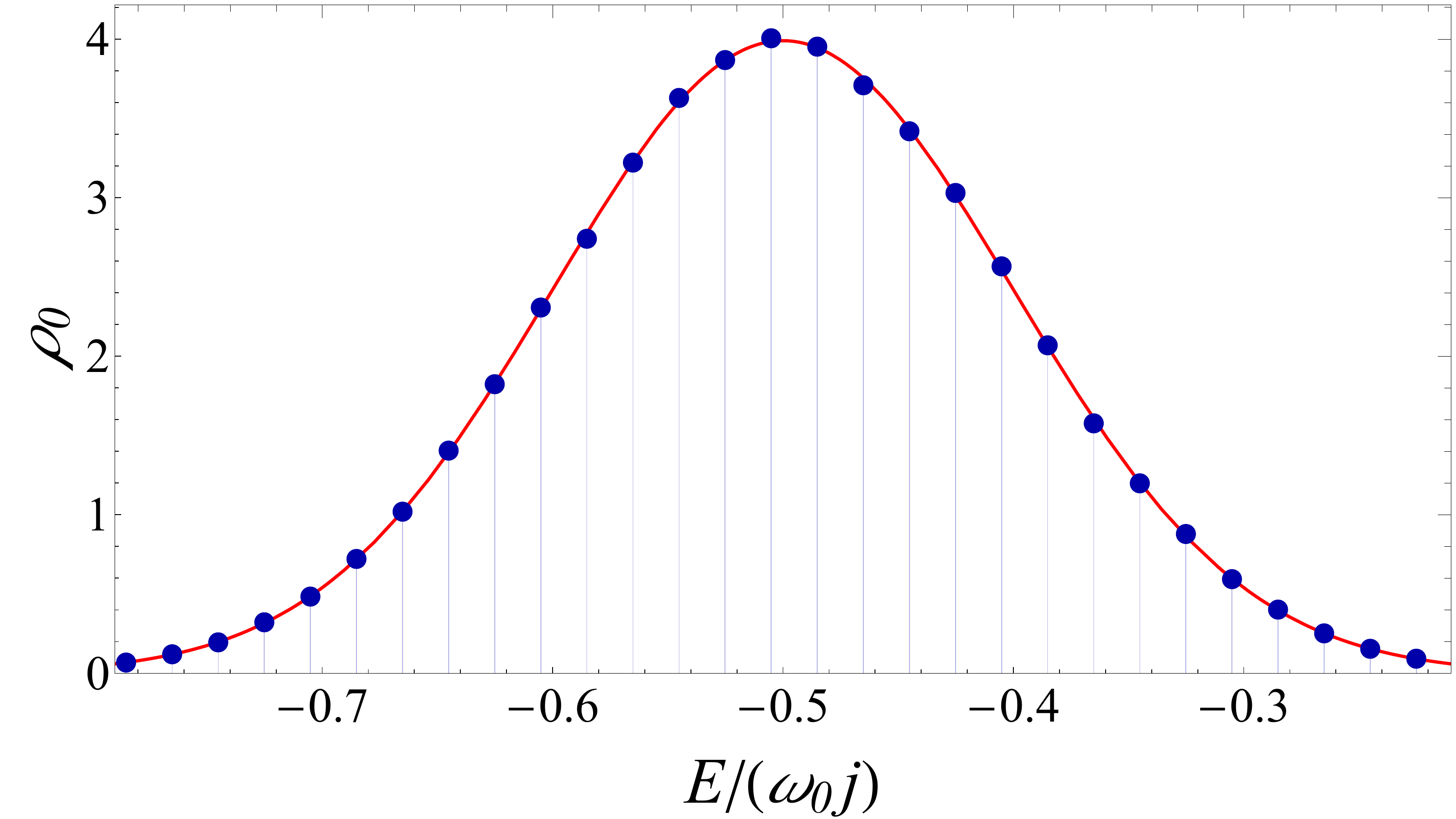}
\end{tabular}
\caption{(Color online) Local density of states (LDoS) for the rectangular (a), strongly bounded  Gaussian (b) and weakly bounded Gaussian (c) energy profiles. Numerical data averaged over 500 random initial states (blue dots) and analytical energy profiles from Eqs.~(\ref{Eq:rect}) and (\ref{Eq:Gauss}) (red solid lines). Bin sizes: $\Delta E=0.01 \omega_0 j$ [(a) and (b)] and  $\Delta E=0.02 \omega_0 j$ (c). The  rectangular LDoS has width $\sigma_R=0.1\omega_0 j$ and the standard deviation of the Gaussian profiles is also $\sigma_G=0.1 \omega_0 j$ with $E_{min}=-0.65 \omega_0 j $ and $E_{max}=-0.35 \omega_0 j$ for (b), whereas for (c) $E_{min}$ is given by the ground-state energy $E_{gs}=-2.125\omega_0 j$ and  by the largest energy obtained for the spectrum $E_{max}=E_T=1.755\omega_0 j$ (both energies are out of the scale used in this panel).}
\label{figS3}
\end{figure*}

In Fig.~\ref{figS3}, we show three cases of energy profiles of the initial state, one rectangular and two Gaussian profiles. The numerical results are obtained by averaging over ensembles of 500 initial states. The agreement with the analytical profiles from Eq.~(\ref{Eq:rect}) and Eq.~(\ref{Eq:Gauss}) confirms that 500 is a sufficiently large number to obtain stable results.

\subsection{Survival Probability: Before the Correlation Hole}

The energy distribution of $|\Psi(0)\rangle $ determines the initial decay of $S_P(t)$. This can be seen by expressing the survival probability in Eq.~(\ref{Eq:SP}) as
$$
S_{P}(t)=\left|\int \rho_{0}(E)e^{-i Et} dE\right|^{2},
$$
where $\rho_{0}(E)=\sum_{k}|c_{k}|^{2}\delta(E-E_{k})$ is the local density of states (LDoS) or strength function, that is the energy distribution weighted by the components of the initial state. If we  approximate the LDoS by its smoothed profile, $\rho_0(E)\approx \rho(E)$,
we obtain for the rectangular profile,
\begin{equation}
S_{P}^{R}(t)= \frac{\sin^2(\sigma_R t)}{(\sigma_R t)^2} ,
\label{eq:SP_R} 
\end{equation}
 and for the  Gaussian profile,
\begin{eqnarray}
&S_{P}^{G}(t)=\dfrac{e^{-\sigma_G^2 t^2}}{{4\cal C}^2} {\cal F}(t), \hspace{0.3 cm} \text{with\ \ \ \ \ \ }
\label{eq:spdecays} 
\\
& {\cal F}(t)\! =\!  \left|\textrm{erf}\left(\dfrac{E_\textrm{c} \! - \! E_\textrm{min} \! - \! i\sigma_G^2 t}{\sqrt{2}\sigma_G}\right) \! - \! \textrm{erf}\left(\dfrac{E_\textrm{c} \! - \! E_\textrm{max} \! - \! i\sigma_G^2 t}{\sqrt{2}\sigma_G}\right)\right|^2.\nonumber
\end{eqnarray}

For very short times, $t\ll \sigma_{R,G}^{-1}$, both $S_{P}^{R}(t)$ and $S_{P}^{G}(t)$ show the universal quadratic decay of the survival probability $1 - \sigma_{R,G}^2 t^2$. For longer times, both profiles lead to a power-law decay $\propto t^{-2}$. This behavior is evident in Eq.~(\ref{eq:SP_R}) and it can be obtained from Eq.~(\ref{eq:spdecays}) by analyzing it at long times, in which case~\cite{Tavora2016,Tavora2017},
\begin{eqnarray}
\label{Eq:resultR1_SM}
&S_{P}^{G}(t\gg\sigma_G^{-1} )\approx \dfrac{1}{2\pi{\cal C}^2\sigma_G^2 t^2}  \times 
\\
&
\left\{
{\cal E}  -2e^{\!-\dfrac{\left[(E_c-E_\textrm{min})^2+ (E_\textrm{max}-E_c)^2
\right]}{2\sigma_G^2}} \hspace{-1.cm} \cos[(E_\textrm{max}-E_\textrm{min})t]
\right\} \!\!,
\nonumber
\end{eqnarray}
where
\begin{equation}
{\cal E}=\exp\left[{-\frac{(E_\textrm{c}-E_\textrm{min})^2}{\sigma_G^2}}\right]+\exp\left[{-\frac{(E_\textrm{c}-E_\textrm{max})^2}{\sigma_G^2}}\right].
\label{Eq:factor}
\end{equation}
Power-law decays of the survival probability are caused by the presence of energy bounds in the LDoS~\cite{Khalfin1958} and the power-law exponent depends on how the bounds are approached~\cite{Urbanowski2009}.

\subsection{Survival Probability: Analytical Expression}

The expressions in Eqs.~(\ref{eq:SP_R}) and (\ref{eq:spdecays}) describe accurately the initial decay of $S_{P}(t)$, for which just the shape and bounds of the envelope of $\rho_{0}(E)$ matters. However, the spectra of finite quantum systems are discrete and, in our case, the eigenvalues are correlated. This results in two additional features to the evolution of $S_{P}(t)$,  beyond the power-law behavior, which are not captured by Eqs.~(\ref{eq:SP_R}) and (\ref{eq:spdecays}). They are the manifestations of the spectrum correlations, which appear at long times, and the saturation of the dynamics to the asymptotic value
\begin{equation}
\label{eqn08}
\bar{S}_{P}=\lim_{t\rightarrow\infty}\frac{1}{t}\int_{0}^{t}dt'S_{P}(t'),
\end{equation}
around which the survival probability oscillates after relaxation.  

To obtain an equation for the full dynamics, we write the survival probability as
\begin{equation}
S_{P}(t)=\sum_{k\neq l}|c_{l}|^{2}|c_{k}|^{2}e^{-i(E_{k}-E_{l})t}+I_{PR} ,
\label{Eq:SP2}
\end{equation}
where
\begin{equation}
I_{PR}\equiv \sum_{k}|\langle\phi_{k}|\Psi (0) \rangle|^{4}=\bar{S}_{P}
\label{Eq:IPR}
\end{equation}
is the so-called inverse participation ratio, which gives the asymptotic temporal value of $S_P(t)$.  The $I_{PR}$ is a measure of the inverse of the number of elements of a given basis (the energy eigenbasis, in our case) participating in an arbitrary quantum state ($|\Psi (0) \rangle$, in our case). 

For the considered ensembles of initial states, it is possible to derive accurate estimates for the ensemble averaged $\langle I_{PR} \rangle$  by using (see Appendix~\ref{App1})
\begin{equation}
\left\langle I_{PR}\right\rangle = \left\langle\frac{\sum_k r_k^2 f^2(E_k)}{\left(\sum_q r_q f(E_q)\right)^2}\right\rangle \approx  \frac{\langle{r_k^2\rangle}}{\langle r_q \rangle^2 } \frac{1}{\nu_c}\int \rho^2(E) dE, 
\label{eq:effdim}
\end{equation}
which, considering  random variables  $r_k$  uniformly  distributed, gives for the rectangular profile,
\begin{equation}
\left\langle I_{PR}^R\right\rangle =\frac{2}{3 \sigma_R \nu_c } ,
\label{eq:iprs}
\end{equation}
and for the Gaussian profile,
\begin{equation}
\left\langle I_{PR}^G\right\rangle =\frac{\textrm{erf}\left(\dfrac{E_\textrm{c}-E_\textrm{min}}{\sigma_G }\right)-\textrm{erf}\left(\dfrac{E_\textrm{c}-E_\textrm{max}}{\sigma_G }\right)}{3\sqrt{\pi}\sigma_G\nu_c{\cal C}^2}.
\label{eq:IPRgauss}
\end{equation}
Above, $\nu_c=\nu(E_c)$ is the DoS evaluated at the central energy $E_c$, which equals
the inverse mean spacing of consecutive energy levels in the region probed by the initial state.

Since the components $c_k$ of the initial state are random numbers, to compute the ensemble average of $S_P(t)$, we can treat the statistical properties of the components and of the spectrum separately. Because the latter has level statistics comparable to that of random matrices from Gaussian orthogonal ensembles (GOE), as shown in Fig.~\ref{figS2}~(a), we can follow steps similar to the ones described in~\cite{Torres2018,SchiulazARXIV} to obtain (see Appendix~\ref{App1} for details), 
\begin{equation}
\left\langle S_{P}(t) \right\rangle =\frac{1- \left\langle I_{PR}\right\rangle }{\eta-1}\left[\eta S_{P}^{bc}(t)-b_{2}\left(\frac{t}{2\pi \nu_c }\right)\right] + \left\langle I_{PR}\right\rangle.
\label{eqn09}
\end{equation}   
Above, $\eta$ is the effective dimension of the ensemble (see Appendix~\ref{App1}) defined as,
\begin{equation}
\eta \equiv \frac{\nu_c}{\int\rho^2(E) dE} =  \frac{\langle r_k^2\rangle}{\langle r_k\rangle^2}\frac{1}{\langle I_{PR}\rangle},
\label{Eq:eta}
\end{equation} 
where the second equality is obtained using Eq.~(\ref{eq:effdim}).
In Eq.~(\ref{eqn09}), $S_{P}^{bc}(t)$ describes the behavior of the survival probability before the manifestations of the correlations between the eigenvalues ($bc$ stands for ``before correlations''), as given by Eqs.~(\ref{eq:SP_R}) and (\ref{eq:spdecays}) for the rectangular and Gaussian profiles. This behavior holds until $\left\langle S_{P}(t) \right\rangle$ reaches its minimum value, which is actually below $\left\langle I_{PR}\right\rangle$. Beyond that, the dynamics becomes controlled by the two-level form factor,
\begin{eqnarray}
b_{2}(\bar{t}) &= & [1-2\bar{t}+\bar{t}\ln(2\bar{t}+1)]\Theta(1-\bar{t})+ \nonumber
\\ 
& +&\left[\bar{t}\ln\left(\frac{2\bar{t}+1}{2\bar{t}-1}\right)-1\right]\Theta(\bar{t}-1),
\label{eqn10}
\end{eqnarray}
where $\Theta$  is the Heaviside step function. The two-level form factor brings the survival probability from its minimum value up to the asymptotic value, creating the dip that is known as correlation hole~\cite{Leviandier1986,Guhr1990,Wilkie1991,Guhr1998,Alhassid1992}. The hole is a direct signature of the presence of correlated eigenvalues, and it does not develop in systems with uncorrelated eigenvalues. The equation used above for $b_{2}(\bar{t})$ is the same used for GOE full random matrices~\cite{MehtaBook}. This implies that beyond the minimum of the correlation hole, the dynamics  shows universal properties. 

The analytical expression for the survival probability in Eq.~(\ref{eqn09}) describes the complete evolution of $\left\langle S_{P}(t) \right\rangle$, from $t=0$ to saturation. The equation has no fitting parameters. All the parameters entering in Eq.~(\ref{eqn09}) can be determined from the properties of the model and the energy profile of the initial state. As we show in the next section, this analytical expression shows remarkable agreement with our numerical results.

%------------------------------------------------
\section{Comparing numerical and analytical results}
\label{sec04}

\begin{figure*}[ht]
\begin{tabular}{cc}
(a)&(b)\\
\includegraphics*[width=.45\textwidth]{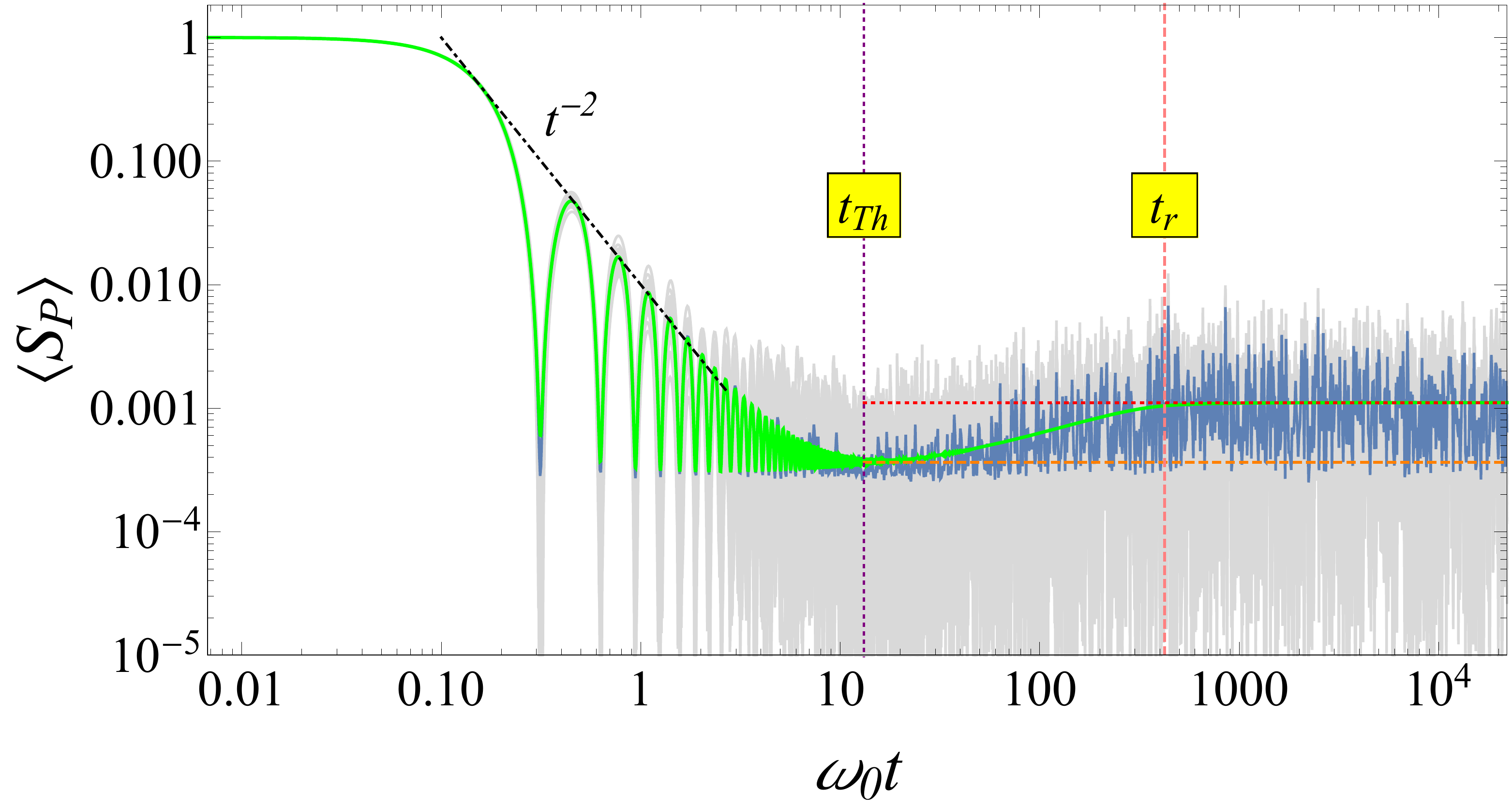}&
\includegraphics*[width=.45\textwidth]{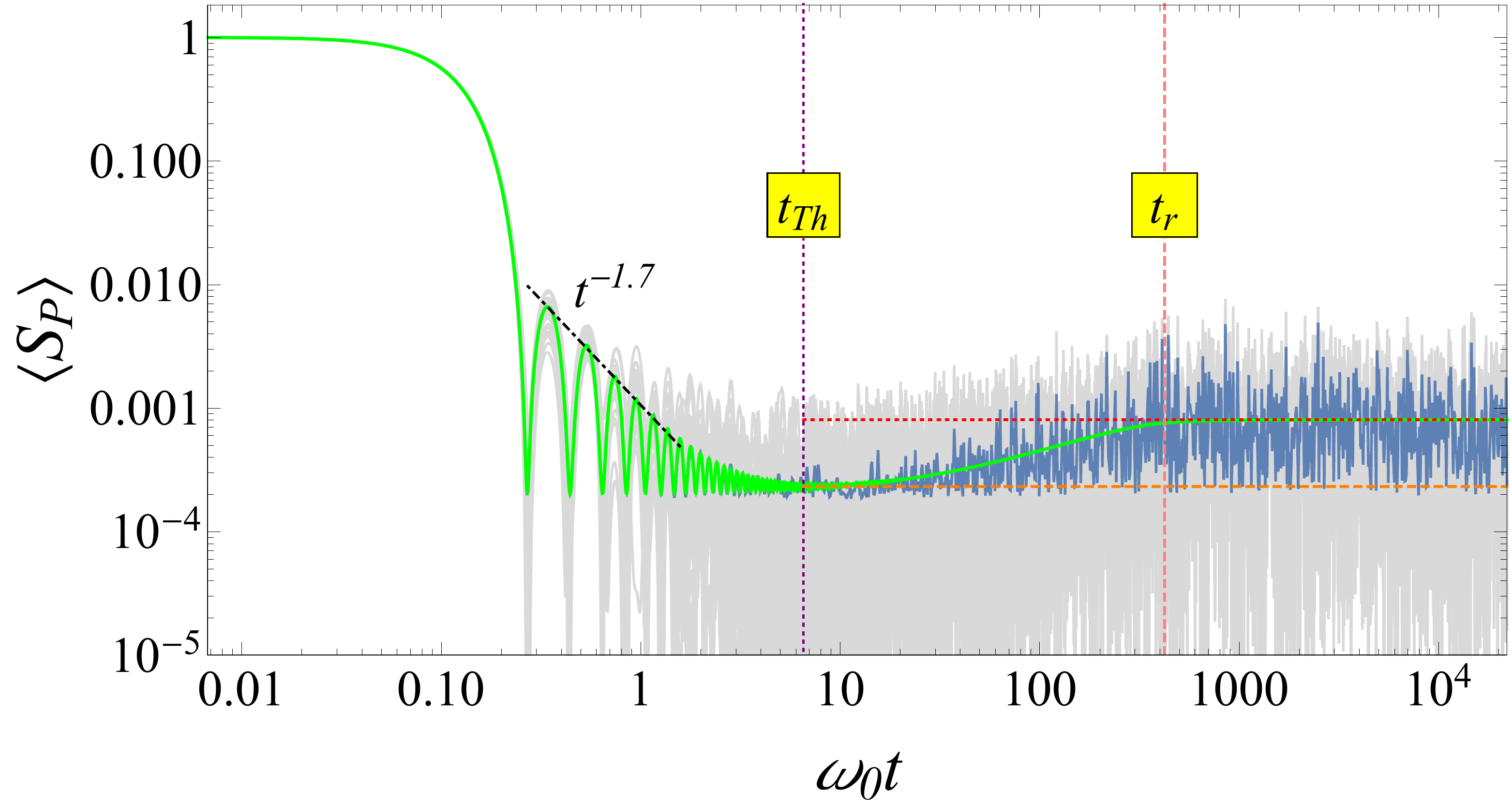}\\
(c)&(d)\\
\includegraphics*[width=0.45\textwidth]{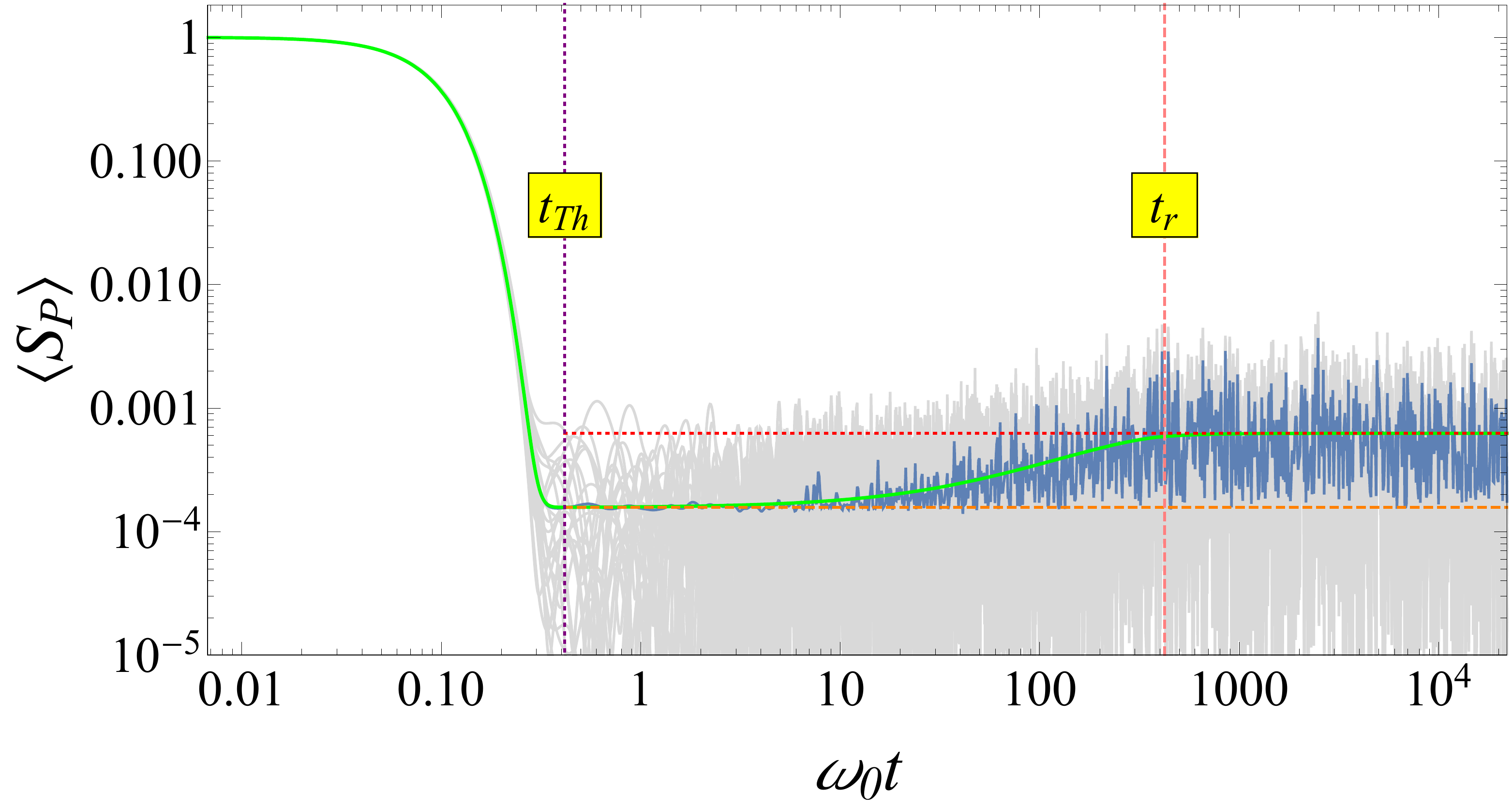}&
\includegraphics*[width=0.45\textwidth]{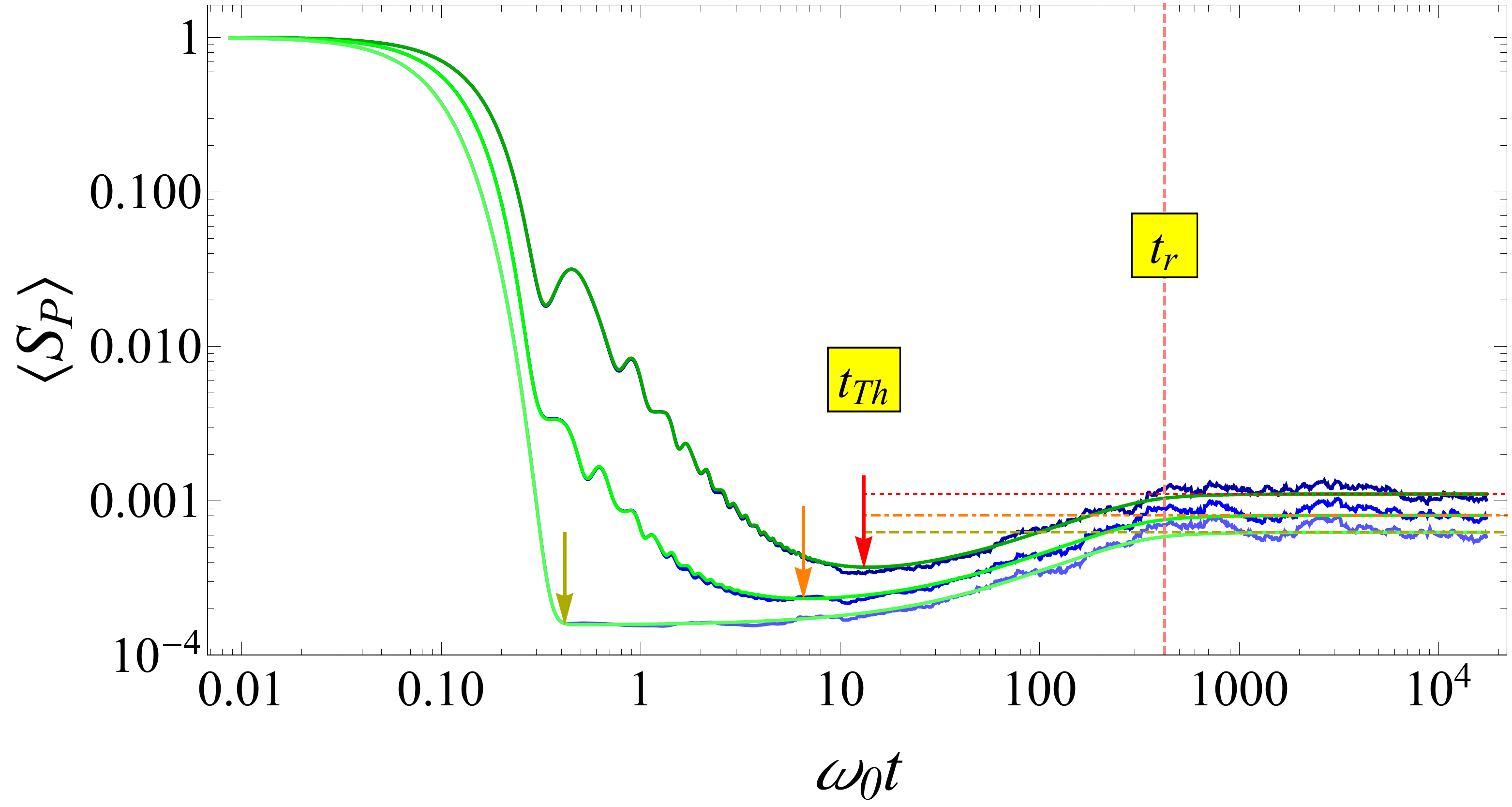}
\end{tabular}
\caption{(Color online) Survival probability for the rectangular (a) [see Fig.~\ref{figS3}~(a)], strongly bounded Gaussian (b) [see Fig.~\ref{figS3}~(b)], and weakly bounded Gaussian (c) [see Fig.~\ref{figS3}~(c)] energy profiles. 
In (a), (b), and (c):
Light (gray) curves depict the survival probability for a single initial random state and the dark (blue) curve represents the ensemble average over 500  random  initial states. The bright (green) line is the analytical expression in Eq.~(\ref{eqn09}). The lowest horizontal dashed line is the analytical estimate for the minimum of the correlation hole (see text). The highest horizontal dotted line shows $\left\langle I_{PR} \right\rangle$, which is $1.106\times10^{-3}$ for (a), $8.032\times10^{-4}$ for (b), and $6.241\times10^{-4}$ for (c). The leftmost vertical line indicates the analytical value for the time when the correlation hole attains its minimum value (Thouless time, $t_{Th}$). 
The rightmost vertical line marks the analytically evaluated relaxation time $t_r$. To determine $t_r$, we fixed $\delta=0.05$ in Eq.~(\ref{Eq:relax}). In (a) and (b), the black dashed line indicates the power law decay $t^{-\lambda}$ for the initial oscillatory decay of $\left\langle S_{P}(t) \right\rangle$. 
Panel (d): The dark (blue) curves represent averages performed overt both initial states and temporal windows of constant size in logarithmic scale. From top to bottom, the curves are obtained with rectangular, strongly bounded Gaussian, and weakly bounded Gaussian energy profiles. The temporal averages are plotted against the mean value of the respective temporal windows. Bright (green) lines depict the same temporal averages of the  analytical expression in Eq.~(\ref{eqn09}).}
\label{fig03}
\end{figure*}

In Fig.~\ref{fig03}, we compare numerical results for the survival probability with the analytical expression given by Eq.~(\ref{eqn09}). The light (gray) lines represent the numerical results obtained with a single initial random state for the rectangular (a) and Gaussian [(b) and (c)] energy profiles. The darker (blue) line is obtained by performing ensemble averages over 500  random initial  states. The bright (green) curve, following extremely well the ensemble average, is the analytical Eq.~(\ref{eqn09}). As clear from all panels, the ensemble average is needed for the hole to be visible.

The initial decay is determined entirely by the energy profile of the initial states, which is the same for every member of the ensemble. An oscillatory decay modulated by a power law $\propto t^{-\lambda}$ is seen in Figs.~\ref{fig03}~(a) and (b). As mentioned before, this behavior is caused by the bounds in the energy profiles, which determine also the size of the oscillations. For the rectangular profile (a), the exponent is indeed $\lambda=2$, as obtained for $S_P^{bc}(t)$ in Eq.~(\ref{eq:SP_R}), whereas for the strongly bounded Gaussian profile (b), the power-law exponent obtained numerically is $\lambda=1.7$, instead of 2 as in Eq.~(\ref{Eq:resultR1_SM}). This is because the oscillations in Fig.~\ref{fig03}~(b) start at a temporal scale where the two-level form factor $b_2$ is not negligible, so it affects the exponent. The power-law decay in Figs.~\ref{fig03}~(a) and (b) is followed by the correlation hole. 
In the case of the weakly bounded Gaussian profile of panel~(c), no trace of the modulated oscillations is left, because they occur at a temporal scale when  $S_P^{bc}(t)$ is already extremely small with respect to the $b_2$ term. In all panels, once saturation is reached,  the survival probability only fluctuates around its asymptotic value.

The minimum of the correlation hole is indicated in Figs.~\ref{fig03}~(a), (b) and (c) with the lowest horizontal dashed line. The dynamics beyond this minimum point depends on level statistics, as confirmed by the fact that the behavior of the ensemble averages is very well described by the analytical expression (\ref{eqn09}), where the same two-level form factor $b_2$ used for full random matrices was employed. However, the size of the temporal fluctuations after the mimimum depends on the fine details of the particular spectrum and on the level of delocalization of the initial state written in the energy eigenbasis~\cite{Reimann2008,Short2011,Short2012,Zangara2013}. 

Contrary to ensembles of random matrices or disordered models, where one can further reduce the temporal fluctuations of $\left\langle S_{P}(t) \right\rangle$ with averages over many energy spectrum realizations, in the case of the Dicke model, the spectrum is exactly the same for every member of the ensemble. Thus, to further reduce the fluctuations in the ensemble averaged $\left\langle S_{P}(t) \right\rangle$, we perform an additional time average over temporal windows of constant size in logarithmic scale, {\em i.e.} temporal windows whose sizes increase exponentially in time. By plotting this temporal averages against the mean time of the respective windows, we obtain the results shown in Figs.~\ref{fig03}~(d). This smoothing procedure results in  numerical curves that are almost identical to the analytical curves, further validating Eq.~(\ref{eqn09}) and the approach that led to it.

The fact that the dynamics beyond the minimum of the correlation hole is governed entirely by the two-level form factor implies that the time to reach saturation depends only on how the $b_2$ function approaches $\left\langle I_{PR} \right\rangle$ (indicated in Fig.~\ref{fig03} with the highest horizontal dashed line). Provided the initial state is fully extended in the energy eigenbasis, counting with the participation of all (most) energy levels in the energy interval characterizing $|\Psi(0)\rangle$, the relaxation time is independent of the initial state. Indeed, as seen in Fig.~\ref{fig03}~(d), the time to reach $\left\langle I_{PR} \right\rangle$, which is shown with the rightmost vertical dashed line, is the same for the three different energy profiles. 
 
\section{Timescales of the survival probability}
\label{ssec02d}

In hands of the analytical expression for the survival probability, we can derive analytically the timescales involved in the relaxation process. We focus on the two longest timescales: the time to reach the minimum of the correlation hole, referred to as Thouless time $t_{Th}$, and the final relaxation time $t_{r}$. 

\subsection{Thouless Time}
The Thouless time divides the dynamics of chaotic systems in two temporal regions, before $t_{Th}$  the dynamics is governed by the shape of the energy distribution of the initial state and the energy bounds, whereas after $t_{Th}$ the dynamics becomes comparable to that obtained with ensembles of full random matrices. The Thouless time marks the point where the term $\eta S_{P}^{bc}(t)$ in Eq.~(\ref{eqn09}) meets the function $b_{2} (t/2\pi \nu_c)$, being derived from
\begin{equation}
\left.\frac{d \left\langle S_{P}(t) \right\rangle }{dt} \right|_{t=t_{Th}}=0 .
\label{eqn11}
\end{equation}
Therefore, we need to examine $S_{P}^{bc}(t)$ at long times and $b_{2} (t/2\pi \nu_c)$ at short times, {\em i.e.} in the temporal range $\sigma^{-1}\ll t\ll \nu_c$ [recall that $\sigma$ is the width of the LDoS and $\nu_c$ is the inverse of the mean level spacing  for the eigenvalues involved in the evolution of $|\Psi(0)\rangle$].

At long times, $\eta S_{P}^{bc}(t)$ shows a power-law decay $\propto t^{-2}$. More specifically, for the rectangular energy profile, the temporal average of the oscillatory decay  in Eq.~(\ref{eq:SP_R})  leads to
\begin{equation}
\eta S_{P}^{bc,R} (t \gg \sigma_R^{-1} ) \rightarrow \frac{\eta}{2 \sigma_R^2 t^2}.
\label{Eq:long_SM_rect}
\end{equation}
For the Gaussian profile, associated with Eq.~(\ref{eq:spdecays}), if the conditions  
\begin{equation}
e^{-(E_{max}-E_c)^2/\sigma^2} >\frac{1}{\eta}  {\hbox {\ \ and\ \ }}   e^{-(E_{min}-E_c)^2/\sigma^2} >\frac{1}{\eta}
\label{Eq:condog}
\end{equation}
are fulfilled, the form of $S_P^{bc}(t)$ in the timescale where it meets $b_2$ is given by [see Eq.~(\ref{Eq:resultR1_SM})]
\begin{equation}
\eta S_{P}^{bc,G}(t\gg\sigma_G^{-1} )\rightarrow\frac{\eta{\cal E}}{2\pi{\cal C}^2\sigma_G^2 t^2} .
\label{Eq:long_SM}
\end{equation}
This is what happens for the strongly bounded Gaussian energy profile of Fig.~\ref{figS3}~(b).
Otherwise, if the conditions in Eq.~(\ref{Eq:condog}) are not fulfilled, the early Gaussian decay still persists at the meeting point with the $b_2$ function and  
\begin{equation}
\eta S_{P}^{bc,G}(t\gg\sigma_G^{-1} )\rightarrow \eta\exp[-\sigma_G^2 t^2].
\label{Eq:long_SM2}
\end{equation}
This is what happens for the weakly bounded Gaussian energy profile of Fig.~\ref{figS3}~(c).

At short times, the two-level form factor is dominated by a linear term, 
\begin{equation}
b_2\left(\frac{t}{2\pi \nu_c }\right) \rightarrow 1 - \frac{t}{\pi \nu_c }  
\hspace{0.4 cm } \text{for} \hspace{0.4 cm } \frac{t}{\nu_c} \ll 1.
\label{Eq:expandb}
\end{equation}

Combining Eqs.~(\ref{Eq:long_SM_rect}) and (\ref{Eq:expandb}), we obtain the Thouless time for the rectangular ensemble   
\begin{equation}
t_{Th}^R=\left(\frac{2\pi\nu_c^2}{\sigma_R}\right)^{1/3},
\label{Eq:tThR}
\end{equation}
where we used that
$\nu_c = \frac{\eta}{2 \sigma_R}$ (see Eq.~(\ref{Eq:eta}) and Appendix~\ref{App1}). 
From Eq.~(\ref{Eq:long_SM}) and Eq.~(\ref{Eq:expandb}), we arrive at the Thouless time for the strongly bounded Gaussian energy profile,
\begin{equation}
t_{Th}^{G}=\left(\frac{\eta\nu_c{\cal E}}{{\cal C}^2\sigma_G^2}\right)^{1/3}
\label{Eq:tThG}
\end{equation}
and if the conditions (\ref{Eq:condog}) are not  fulfilled, using Eq.~(\ref{Eq:long_SM2}), we have
\begin{equation}
t_{Th}^{G}\approx \frac{\sqrt{\log(2\pi\eta\sigma_G \nu_c)}}{\sigma}
\label{Eq:tThG2}.
\end{equation}
The Thouless time obtained in the equations above are indicated in Fig.~\ref{fig03},  showing excellent agreement with the numerics.

\subsection{Relaxation Time}
The relaxation time depends only on the $b_2$ function at long times, which grows toward saturation following a power-law behavior,
\begin{equation}
b_2\left( \frac{t}{2\pi \nu_c } \right) \rightarrow  \frac{\pi^2 \nu_c^2}{3 t^2} \hspace{0.4 cm } \text{for} \hspace{0.4 cm } \frac{t}{\nu_c} \gg 1 .
\label{Eq:longB2}
\end{equation}
Even though Eq.~(\ref{Eq:long_SM_rect}) and Eq.~(\ref{Eq:long_SM}) decay with the same power-law exponent 2, as in Eq.~(\ref{Eq:longB2}), the latter is proportional to $\eta^2$, since $\nu_c \propto \eta$, while the former equations are proportional to $\eta$, which justifies considering only Eq.~(\ref{Eq:longB2}).
We define the relaxation time according to
\begin{equation}
\left\langle S_{P}(t_r) \right\rangle = (1-\delta) \left\langle I_{PR} \right\rangle ,
\end{equation}
where $\delta$ is a small parameter determining the point where $\left\langle S_{P}(t) \right\rangle$ is already within the fluctuations around the asymptotic value. We arrive at
\begin{equation}
t_{r}=\frac{\pi\nu_c}{2\sqrt{\delta}} ,
\label{Eq:relax}
\end{equation}
which holds for the three  energy profiles. This time is proportional to the inverse of the mean level spacing, $\nu_c$, which is the largest timescale of a quantum system and is known as the Heisenberg time. In  Fig.~\ref{fig03}, $t_{r}$ is indicated with the rightmost vertical lines, showing a excellent agreement with the numerical results.
 
\subsection{Scaling with system size: Thouless and relaxation time}

With Eqs.~(\ref{Eq:tThR}), (\ref{Eq:tThG}), (\ref{Eq:tThG2}) and (\ref{Eq:relax}), we can determine the dependence of the Thouless and relaxation times on the size of the system, {\em i.e}  on the number $\mathcal{N}=2j$ of two-level atoms. Both times depend on $\nu_c$, which scales linearly with $j$. We can write $\nu_c= \nu_o j/\omega$ and evaluate $\nu_o$ numerically, which for the chosen energy $E_c$  is $\nu_o=0.6027$.

The Thouless time depends additionally on the widths $\sigma_R$ and $\sigma_G$  of the energy distribution of the initial state and, for the Gaussian profile, on the energy bounds $E_{min}$ and $E_{max}$. The scaling $\sigma_{R,G}=\sigma_{R,G}^o j^\beta$ of these widths, as well as the scalings of $(E_c-E_{min})\propto j^{\alpha_1}$ and $(E_{max}-E_c)\propto j^{\alpha_2}$, can in principle be selected at will in the range $-1\leq \beta\leq 1$ and $-1\leq \alpha_i\leq 1$.  The lower values $\beta=\alpha_i=-1$ are imposed by the scaling of the mean-level spacing of consecutive energy levels and the upper bound is given by the scaling of the energy spectrum, which is proportional to $j$.
A physical relevant choice for the previous scalings is $\beta=1/2$, which is the scaling of the energy  widths of minimal uncertainty coherent states \cite{Schliemann2015}, and $\alpha_i=1$, which implies that the bounding energy interval of the Gaussian profile scales as the energy spectrum. Therefore for the rectangular profile and strongly bounded Gaussian profile satisfying  conditions (\ref{Eq:condog}), the Thouless time scales as
\begin{equation}
t_{Th}^{R,G}= t_o^{R,G} j^{1/2},
\end{equation}
where $t_o^{R,G}$ is a constant determined by $\nu_o$, and $\sigma_{R,G}^o$. For the  rectangular profile, this scaling is valid for any $j$, but for the Gaussian profile, it is valid up to a finite value of $j$.   This is because we assume that  $(E_c-E_{min})$ and  $(E_{max}-E_c)$ grow with $j$ faster than $\sigma_G$, which implies that for large enough $j$  the conditions (\ref{Eq:condog}) will not be satisfied anymore, switching to the scenario of Eq.~(\ref{Eq:tThG2}). For the weakly bounded Gaussian profile,  described by Eq.~(\ref{Eq:tThG2}),  the Thouless time  for large $j$ is given by,  
\begin{equation}
t_{Th}^G= \frac{\sqrt{\log c_o+3\log j}}{\sigma_G^o j^{1/2}},
\label{eq:thg}
\end{equation}
where $c_o=4 \pi^{(3/2)} (\sigma_G^o)^2 \nu_o^2/\omega^2$, and we have approximated the error functions by their asymptotic values, $\lim_{z\rightarrow\infty}\textrm{erf}(z)\rightarrow 1$.
 
The relaxation time is independent of the details of the initial state and scales linearly with $\nu_c$,  so it is given simply by
\begin{equation}
t_{r}=\frac{\pi \nu_o}{2\omega\sqrt{\delta}} j.
\end{equation}
The distance between the Thouless and the relaxation time diverge with $j$, which means that the correlation hole gets elongated as the number of atoms increases. 

\subsection{Depth of the correlation hole}

With the Thouless time, we can quantify the relative depth of the correlation hole through the expression
\begin{equation}
\kappa=\frac{\left\langle I_{PR} \right\rangle  - \left\langle S_{P} (t_{Th}) \right\rangle }{\left\langle I_{PR} \right\rangle}.
\label{eqn12}
\end{equation}
For the ensembles considered in this paper, we can calculate the depth of the correlation hole for $j\gg 1$.   A direct substitution of the Thouless time in the analytical expression for the survival probability in Eq.~(\ref{eqn09}) allows to demonstrate that
$$
\lim_{j\gg 1 } \left\langle S_{P} (t_{Th}) \right\rangle = -\frac{1}{\eta}+\left\langle I_{PR} \right\rangle = \frac{\langle r_k^2\rangle-\langle r_k\rangle^2}{\langle r_k^2\rangle} \left\langle I_{PR} \right\rangle,
$$ 
where in the last step we have  used Eq.~(\ref{Eq:eta}). From this result, we obtain 
\begin{equation}
\kappa_\infty=\frac{\langle r_k\rangle^2}{\langle r_k^2\rangle}. 
\label{Eq:kappin}
\end{equation}
The value $\kappa_\infty$ gives an upper bound for the depth of the correlation hole for finite $j$. In the case of random variables $r_k$ uniformly distributed in the interval $[0,1]$, as considered here~\cite{noteGOE}, this bound is  $\kappa_\infty=3/4$. 

The actual values of $\kappa$ for the finite systems studied, where $j=100$, are obtained by substituting Eqs.~(\ref{Eq:tThR}), (\ref{Eq:tThG}) and (\ref{Eq:tThG2}) in Eq. (\ref{eqn09}), which gives $ \left\langle S_{P} (t_{Th}) \right\rangle$, and by getting $\left\langle I_{PR} \right\rangle$ from Eqs.~(\ref{eq:iprs}) and (\ref{eq:IPRgauss}).  We get $\kappa=0.672$ for the rectangular ensemble, $\kappa=0.711$ for the ensemble from the strongly bounded Gaussian profile, and $\kappa=0.748$ for the ensemble from the weakly bounded Gaussian profile.  The analytical estimates for $ \left\langle S_{P} (t_{Th}) \right\rangle$ and $\left\langle I_{PR} \right\rangle$ are depicted in Fig.~\ref{fig03}~(a), (b), and (c) with horizontal lines, showing excellent  agreement with the numerical results. This confirms that the analytical expression in Eq.~(\ref{eqn09}) describes the survival probability at any timescale.

\section{CONCLUSIONS}
\label{sec06}

We obtained an analytical expression that describes remarkably well the entire evolution of the averaged survival probability, $\langle S_P (t) \rangle$, for the Dicke model in the chaotic regime and allowed us to derive analytical expressions for the different timescales involved in the relaxation to equilibrium. Due to spectral correlations, the survival probability exhibits a correlation hole. We find that the initial decay of $\langle S_P (t) \rangle$  and the time $t_{Th}$ for it to reach the minimum of the correlation hole  (Thouless time) depend on the energy profile of the initial states. %
Beyond the Thouless time, the dynamics is universal, being governed by the two-level form factor of the GOE. This implies that the time, $t_r$, for the survival probability to reach equilibrium (relaxation time) depends only on the energies, being proportional to the inverse of the mean level spacing. %
An interesting future direction would be to investigate the timescales involved in the relaxation process of other physical observables that are relevant for the Dicke model.

\section*{Acknowledgements}
We acknowledge the support of the Computation Center- ICN, in particular to Enrique Palacios,  Luciano Diaz and Eduardo Murrieta. D.V, J.G.H, M.A.B.-M. and S.L.-H acknowledge Jorge Ch\'avez-Carlos for fruitful discussions and his technical support.  E.J.T.-H. acknowledges funding from VIEP-BUAP (Grant Nos. MEBJ-EXC19-G, LUAG- EXC19-G), Mexico.  L.F.S. is supported by the NSF Grant No.~DMR-1603418. S.L.-H. acknowledges financial support from Mexican CONACyT project CB2015-01/255702, J.G.H. and D.V. acknowledge funding from DGAPA- UNAM project IN109417. 

\appendix
\section{Ensemble averages of the survival probability}
\label{App1}

In this appendix we present the steps involved in the derivation of the analytical expression given by Eq.~(\ref{eqn09}) for the ensemble average of the survival probability. We begin with Eq.~(\ref{Eq:SP2}) and perform ensemble averages, taking into account that the eigenvalues and the components $|c_{k}|^{2}$ of the initial state are statistically independent,
\begin{eqnarray}
\langle S_{P}(t) \rangle &=&\left\langle \sum_{k\neq l}|c_{l}|^{2}|c_{k}|^{2}e^{-i(E_{k}-E_{l})t}\right\rangle+\left\langle I_{PR}\right\rangle
\nonumber\\
&=& \sum_{k\neq l} \left\langle|c_{l}|^{2}|c_{k}|^{2}\right\rangle e^{-i(E_{k}-E_{l})t}+\left\langle I_{PR}\right\rangle.\label{eq:ap0}
\end{eqnarray}

Let us consider first the ensemble average of $I_{PR}$. Using Eq.~(\ref{eq:components}), we have 
\begin{eqnarray}
\left\langle I_{PR}\right\rangle&=&\left\langle \sum_k |c_k|^4\right\rangle= \left\langle\frac{\sum_k r_k^2 f^2(E_k)}{\left(\sum_q r_q f(E_q)\right)^2}\right\rangle
\nonumber\\
&=&\sum_k\left\langle\frac{ r_k^2 }{\left(\sum_q r_q f(E_q)\right)^2}\right\rangle f^2(E_k).\label{eq:ap1}
\end{eqnarray}
For a  large number of  components, the average of the second line above  can be approximated as
\begin{eqnarray}
\left\langle \frac{r_k^2}{\left(\sum_q r_q f(E_q)\right)^2}\right\rangle&\approx&  \frac{\langle r_k^2\rangle}{\left(\sum_q \langle r_q \rangle f(E_q)\right)^2}\nonumber\\&=&%\frac{4}{3}
\frac{\langle r_k^2\rangle}{\langle r_q\rangle^2}\frac{1}{\left(\sum_{q} f(E_q)\right)^2},
\end{eqnarray}
where in the last equality we have used  the fact that $\langle r_q^n\rangle $ is actually independent of index $q$.
 By inserting this result in Eq.~(\ref{eq:ap1}), we obtain
\begin{equation}
\left\langle I_{PR}\right\rangle=%\frac{4}{3}
\frac{\langle r_k^2\rangle}{\langle r_q\rangle^2}\frac{\sum_k f^2(E_k)}{\left(\sum_{q} f(E_q)\right)^2}\equiv %\frac{4}{3\eta}
\frac{\langle r_k^2\rangle}{\langle r_q\rangle^2}\frac{1}{\eta}.
\label{eq:ap1b}
\end{equation}
Here, we have introduced the effective dimension of the ensemble 
\begin{equation}
\eta=\frac{\left(\sum_{q} f(E_q)\right)^2}{\sum_k f^2(E_k)},
\label{eq:apeta}
\end{equation}
whose name comes from the fact that it reduces to the number of states participating in the rectangular ensemble, as it is shown below.
We now approximate the sums in Eq.~(\ref{eq:ap1b}) by  integrals,
\begin{equation}
\sum_k \bullet\rightarrow \int dE\, \nu(E) \bullet ,
\label{eq:ap2}
\end{equation} 
to   obtain  
$$
\left\langle I_{PR}\right\rangle\approx %\frac{4}{3} 
\frac{\langle r_k^2\rangle}{\langle r_q\rangle^2}
\frac{\int dE\, \rho^2(E)/\nu(E)  }{\left(\int dE\, \rho(E)\right)^2 }=\frac{\langle r_k^2\rangle}{\langle r_q\rangle^2}\int dE\, \frac{\rho^2(E)}{\nu(E)},
$$
where we have used $f(E)=\rho(E)/\nu(E)$, and, in the last equality, the normalization of $\rho(E)$. Finally, since $\nu(E)$ varies linearly in the energy interval where $\rho(E)$ is significant, we substitute the function $\nu(E)$ by its value in the center of the profile distribution $\nu_c\equiv\nu(E_c)$ and obtain
 $$
\left\langle I_{PR}\right\rangle \approx 
\frac{\langle r_k^2\rangle}{\langle r_q\rangle^2}\frac{1}{\nu_c}
%\frac{4}{3\nu_c}
\int dE\, \rho^2(E),
$$
and 
$$\eta=\frac{\nu_c}{\int dE\, \rho^2(E)}.$$
From the expression for $\eta$,  it is clear that, in the case of the rectangular profile $\eta= 2\nu_c\sigma_R $, which  gives the number of states participating in the ensemble.  

For the first term in Eq~(\ref{eq:ap0}), we have to evaluate the ensemble average 
$$
 \left\langle |c_l|^2 |c_k|^2\right\rangle= f_l f_k \left\langle \frac{r_l r_k}{\left(\sum_q r_q f_q\right)^2} \right\rangle 
$$
where $(l\not= k)$ and, for simplicity,  we introduced the shorthand notation $f_k\equiv f(E_k)$.
To obtain an approximation to the average, we consider the identity,
$$
1=\frac{\sum_k f_k r_k\sum_{l}f_l r_l}{\left(\sum_q r_q f_q\right)^2}=\frac{\sum_k f_k^2 r_k^2}{\left(\sum_q r_q f_q\right)^2}+\frac{\sum_{l\not=k} r_l r_k f_l f_k}{\left(\sum_q r_q f_q\right)^2}.
$$
From this, we obtain 
$$
\frac{\sum_{l\not=k} r_l r_k f_l f_k}{\left(\sum_q r_q f_q\right)^2}=1-I_{PR}.
$$
By taking the ensemble average of this expression and assuming that  $$\left\langle \frac{r_l r_k}{\left(\sum_q r_q f_q\right)^2} \right\rangle\approx A,$$ where $A$ is a constant independent of indexes $l$ and $k$, we get
$$
A \sum_{l\not=k} f_l f_k=1-\langle{I}_{PR}\rangle, 
$$
which implies that
$$ 
 A=\frac{1-\langle{I}_{PR}\rangle}{\sum_{l\not=k} f_l f_k }.
$$
 The sum in the denominator of the  equation above can be expressed in terms of the effective dimension $\eta$ [see Eq.~(\ref{eq:apeta})], as follows
$$
\sum_{l\not=k} f_l f_k=\frac{\eta-1}{\eta}\left(\sum_q f_q\right)^2=\frac{\eta-1}{\eta},
$$
where in the last step, the normalization $\sum_q f_q\rightarrow \int dE \rho(E)=1$ was used. With the the above result,  the average  can be written as
$$
 \left\langle |c_l|^2 |c_k|^2\right\rangle\approx\frac{1-\langle I_{PR}\rangle}{\eta-1}\eta f_l f_k, 
$$
which, when substituted in Eq.~(\ref{eq:ap0}),  leads to
\begin{equation}
\langle S_{P}(t) \rangle= \langle I_{PR}\rangle+\frac{1-\langle I_{PR}\rangle}{\eta-1} \eta\sum_{k\neq l} f_k f_l    e^{-i(E_{k}-E_{l})t}.
\label{eq:apSP0}
\end{equation}

We now turn our attention to the double sum $\sum_{k\neq l}$ in Eq.~(\ref{eq:apSP0}). To solve it, we use
$$
\sum_{k\not=l}  \bullet \rightarrow \int dE dE'  R(E,E')   \bullet ,
$$
where the Dyson two-point correlation function,
$R(E,E')=\nu(E)\nu(E')-T(E-E')$, includes the DoS, $\nu(E)$, and  the two-level cluster function, $T(E-E')$, which has information about the correlations between the eigenvalues~\cite{MehtaBook}. We then obtain
\begin{eqnarray}
\sum_{k\neq l} f_k f_l  e^{-i(E_{k}-E_{l})t}\rightarrow \left|\int \rho(E) e^{-i E t} dE\right|^2 
\nonumber
\\
-\int dE dE' \rho(E)\rho(E') \frac{T(E-E')}{\nu(E)\nu(E')} e^{-i(E-E')t}.
\label{Eq:yy}
\end{eqnarray}
Using unfolded energy variables, which leads to universal functions in the limit of an infinite number of levels, we have~\cite{MehtaBook}
$$
Y\left([E-E']\nu_c\right)= T(E-E')/\nu_c^2.
$$  
With this function, the second integral in Eq.~(\ref{Eq:yy}) is
\begin{eqnarray}\int dE dE' \rho(E)\rho(E') \frac{T(E-E')}{\nu(E)\nu(E')} e^{-i(E-E')t}\approx
\nonumber \\
 \int dE dE'  \rho(E)\rho(E')Y\left([E-E']\nu_c\right) e^{-i(E-E')t},
\end{eqnarray}
which, in terms of variables $z=E'$ and $x=(E-E')\nu_c$ becomes
\begin{equation}\frac{1}{\nu_c} \int dx dz \rho(z)\rho(z+x/\nu_c) Y(x)e^{-i 2\pi x\tilde{t}}
\end{equation}
with $\tilde{t}= t/(2 \pi \nu_c)$. By expanding  $\rho(z+x/\nu_c)$ in powers of $x$ and considering only the lowest order, the double integral can be approximated by a product of two independent integrals
$$
\frac{1}{\nu_c} \int dz \rho(z)^2 \int dx Y(x)e^{-i 2\pi x\tilde{t}}=\frac{1}{\eta}b_2\left(\frac{t}{2\pi \nu_c}\right),
$$
where we have used  the effective dimension introduced before, and $b_2(\tilde{t})$ is the known Fourier transform of the GOE-two level cluster function~\cite{MehtaBook},  the so called  two-level form factor  shown in  Eq.~(\ref{eqn10}). We use the same $b_2(\tilde{t})$ as in GOE matrices, because the unfolded spectrum of the Dicke model has correlations comparable to those of the GOE levels. 

Gathering the previous results together, we  obtain for the ensemble average of the survival probability 
\begin{eqnarray}
\langle S_{P}(t) \rangle  &=& \langle I_{PR}\rangle +\\
\frac{ 1-\langle I_{PR}\rangle}{\eta-1}& &\left[\eta \left|
\int \rho(E) e^{-i E t} dE
\right|^2 - b_2\left(\frac{t}{2\pi \nu_c}\right)\right].\nonumber
\end{eqnarray}
Since the  squared absolute value inside the parenthesis is  the short time decay  $S_p^{bc}$ given by Eqs.~(\ref{eq:SP_R}) and (\ref{eq:spdecays}), we finally arrive at our analytical expression in  Eq.~(\ref{eqn09}).

%----------------------------------------------------------------------------------------
%	REFERENCE LIST
%----------------------------------------------------------------------------------------
%

%----------------------------------------------------------------------------------------


\begin{thebibliography}{79}%
\makeatletter
\providecommand \@ifxundefined [1]{%
 \@ifx{#1\undefined}
}%
\providecommand \@ifnum [1]{%
 \ifnum #1\expandafter \@firstoftwo
 \else \expandafter \@secondoftwo
 \fi
}%
\providecommand \@ifx [1]{%
 \ifx #1\expandafter \@firstoftwo
 \else \expandafter \@secondoftwo
 \fi
}%
\providecommand \natexlab [1]{#1}%
\providecommand \enquote  [1]{``#1''}%
\providecommand \bibnamefont  [1]{#1}%
\providecommand \bibfnamefont [1]{#1}%
\providecommand \citenamefont [1]{#1}%
\providecommand \href@noop [0]{\@secondoftwo}%
\providecommand \href [0]{\begingroup \@sanitize@url \@href}%
\providecommand \@href[1]{\@@startlink{#1}\@@href}%
\providecommand \@@href[1]{\endgroup#1\@@endlink}%
\providecommand \@sanitize@url [0]{\catcode `\\12\catcode `\$12\catcode
  `\&12\catcode `\#12\catcode `\^12\catcode `\_12\catcode `\%12\relax}%
\providecommand \@@startlink[1]{}%
\providecommand \@@endlink[0]{}%
\providecommand \url  [0]{\begingroup\@sanitize@url \@url }%
\providecommand \@url [1]{\endgroup\@href {#1}{\urlprefix }}%
\providecommand \urlprefix  [0]{URL }%
\providecommand \Eprint [0]{\href }%
\providecommand \doibase [0]{http://dx.doi.org/}%
\providecommand \selectlanguage [0]{\@gobble}%
\providecommand \bibinfo  [0]{\@secondoftwo}%
\providecommand \bibfield  [0]{\@secondoftwo}%
\providecommand \translation [1]{[#1]}%
\providecommand \BibitemOpen [0]{}%
\providecommand \bibitemStop [0]{}%
\providecommand \bibitemNoStop [0]{.\EOS\space}%
\providecommand \EOS [0]{\spacefactor3000\relax}%
\providecommand \BibitemShut  [1]{\csname bibitem#1\endcsname}%
\let\auto@bib@innerbib\@empty
%</preamble>
\bibitem [{\citenamefont {Reimann}(2008)}]{Reimann2008}%
  \BibitemOpen
  \bibfield  {author} {\bibinfo {author} {\bibfnamefont {Peter}\ \bibnamefont
  {Reimann}},\ }\bibfield  {title} {\enquote {\bibinfo {title} {Foundation of
  statistical mechanics under experimentally realistic conditions},}\
  }\href@noop {} {\bibfield  {journal} {\bibinfo  {journal} {Phys. Rev. Lett.}\
  }\textbf {\bibinfo {volume} {101}},\ \bibinfo {pages} {190403} (\bibinfo
  {year} {2008})}\BibitemShut {NoStop}%
\bibitem [{\citenamefont {Short}(2011)}]{Short2011}%
  \BibitemOpen
  \bibfield  {author} {\bibinfo {author} {\bibfnamefont {A.~J.}\ \bibnamefont
  {Short}},\ }\bibfield  {title} {\enquote {\bibinfo {title} {Equilibration of
  quantum systems and subsystems},}\ }\href@noop {} {\bibfield  {journal}
  {\bibinfo  {journal} {New J. Phys.}\ }\textbf {\bibinfo {volume} {13}},\
  \bibinfo {pages} {053009} (\bibinfo {year} {2011})}\BibitemShut {NoStop}%
\bibitem [{\citenamefont {Short}\ and\ \citenamefont
  {Farrelly}(2012)}]{Short2012}%
  \BibitemOpen
  \bibfield  {author} {\bibinfo {author} {\bibfnamefont {A.~J.}\ \bibnamefont
  {Short}}\ and\ \bibinfo {author} {\bibfnamefont {T.~C.}\ \bibnamefont
  {Farrelly}},\ }\bibfield  {title} {\enquote {\bibinfo {title} {Quantum
  equilibration in finite time},}\ }\href@noop {} {\bibfield  {journal}
  {\bibinfo  {journal} {New J. Phys.}\ }\textbf {\bibinfo {volume} {14}},\
  \bibinfo {pages} {013063} (\bibinfo {year} {2012})}\BibitemShut {NoStop}%
\bibitem [{\citenamefont {Zangara}\ \emph {et~al.}(2013)\citenamefont
  {Zangara}, \citenamefont {Dente}, \citenamefont {Torres-Herrera},
  \citenamefont {Pastawski}, \citenamefont {Iucci},\ and\ \citenamefont
  {Santos}}]{Zangara2013}%
  \BibitemOpen
  \bibfield  {author} {\bibinfo {author} {\bibfnamefont {Pablo~R.}\
  \bibnamefont {Zangara}}, \bibinfo {author} {\bibfnamefont {Axel~D.}\
  \bibnamefont {Dente}}, \bibinfo {author} {\bibfnamefont {E.~J.}\ \bibnamefont
  {Torres-Herrera}}, \bibinfo {author} {\bibfnamefont {Horacio~M.}\
  \bibnamefont {Pastawski}}, \bibinfo {author} {\bibfnamefont {A.}~\bibnamefont
  {Iucci}}, \ and\ \bibinfo {author} {\bibfnamefont {Lea~F.}\ \bibnamefont
  {Santos}},\ }\bibfield  {title} {\enquote {\bibinfo {title} {Time
  fluctuations in isolated quantum systems of interacting particles},}\
  }\href@noop {} {\bibfield  {journal} {\bibinfo  {journal} {Phys. Rev. E}\
  }\textbf {\bibinfo {volume} {88}},\ \bibinfo {pages} {032913} (\bibinfo
  {year} {2013})}\BibitemShut {NoStop}%
\bibitem [{\citenamefont {He}\ \emph {et~al.}(2013)\citenamefont {He},
  \citenamefont {Santos}, \citenamefont {Wright},\ and\ \citenamefont
  {Rigol}}]{HeSantos2013}%
  \BibitemOpen
  \bibfield  {author} {\bibinfo {author} {\bibfnamefont {K.}~\bibnamefont
  {He}}, \bibinfo {author} {\bibfnamefont {L.~F.}\ \bibnamefont {Santos}},
  \bibinfo {author} {\bibfnamefont {T.~M.}\ \bibnamefont {Wright}}, \ and\
  \bibinfo {author} {\bibfnamefont {M.}~\bibnamefont {Rigol}},\ }\bibfield
  {title} {\enquote {\bibinfo {title} {Single-particle and many-body analyses
  of a quasiperiodic integrable system after a quench},}\ }\href {\doibase
  10.1103/PhysRevA.87.063637} {\bibfield  {journal} {\bibinfo  {journal} {Phys.
  Rev. A}\ }\textbf {\bibinfo {volume} {87}},\ \bibinfo {pages} {063637}
  (\bibinfo {year} {2013})}\BibitemShut {NoStop}%
\bibitem [{\citenamefont {Gogolin}\ and\ \citenamefont
  {Eisert}(2016)}]{Gogolin2016}%
  \BibitemOpen
  \bibfield  {author} {\bibinfo {author} {\bibfnamefont {C.}~\bibnamefont
  {Gogolin}}\ and\ \bibinfo {author} {\bibfnamefont {J.}~\bibnamefont
  {Eisert}},\ }\bibfield  {title} {\enquote {\bibinfo {title} {Equilibration,
  thermalisation, and the emergence of statistical mechanics in closed quantum
  systems},}\ }\href {http://stacks.iop.org/0034-4885/79/i=5/a=056001}
  {\bibfield  {journal} {\bibinfo  {journal} {Rep. Prog. Phys.}\ }\textbf
  {\bibinfo {volume} {79}},\ \bibinfo {pages} {056001} (\bibinfo {year}
  {2016})}\BibitemShut {NoStop}%
\bibitem [{\citenamefont {Borgonovi}\ \emph {et~al.}(2016)\citenamefont
  {Borgonovi}, \citenamefont {Izrailev}, \citenamefont {Santos},\ and\
  \citenamefont {Zelevinsky}}]{Borgonovi2016}%
  \BibitemOpen
  \bibfield  {author} {\bibinfo {author} {\bibfnamefont {F.}~\bibnamefont
  {Borgonovi}}, \bibinfo {author} {\bibfnamefont {F.~M.}\ \bibnamefont
  {Izrailev}}, \bibinfo {author} {\bibfnamefont {L.~F.}\ \bibnamefont
  {Santos}}, \ and\ \bibinfo {author} {\bibfnamefont {V.~G.}\ \bibnamefont
  {Zelevinsky}},\ }\bibfield  {title} {\enquote {\bibinfo {title} {Quantum
  chaos and thermalization in isolated systems of interacting particles},}\
  }\href {\doibase 10.1016/j.physrep.2016.02.005} {\bibfield  {journal}
  {\bibinfo  {journal} {Phys. Rep.}\ }\textbf {\bibinfo {volume} {626}},\
  \bibinfo {pages} {1} (\bibinfo {year} {2016})}\BibitemShut {NoStop}%
\bibitem [{\citenamefont {Alessio}\ \emph {et~al.}(2016)\citenamefont
  {Alessio}, \citenamefont {Kafri}, \citenamefont {Polkovnikov},\ and\
  \citenamefont {Rigol}}]{Dalessio2016}%
  \BibitemOpen
  \bibfield  {author} {\bibinfo {author} {\bibfnamefont {L.~D'}\ \bibnamefont
  {Alessio}}, \bibinfo {author} {\bibfnamefont {Y.}~\bibnamefont {Kafri}},
  \bibinfo {author} {\bibfnamefont {A.}~\bibnamefont {Polkovnikov}}, \ and\
  \bibinfo {author} {\bibfnamefont {M.}~\bibnamefont {Rigol}},\ }\bibfield
  {title} {\enquote {\bibinfo {title} {From quantum chaos and eigenstate
  thermalization to statistical mechanics and thermodynamics},}\ }\href
  {\doibase 10.1080/00018732.2016.1198134} {\bibfield  {journal} {\bibinfo
  {journal} {Adv. Phys.}\ }\textbf {\bibinfo {volume} {65}},\ \bibinfo {pages}
  {239--362} (\bibinfo {year} {2016})}\BibitemShut {NoStop}%
\bibitem [{\citenamefont {Dymarsky}({\natexlab{a}})}]{DymarskyARXIV}%
  \BibitemOpen
  \bibfield  {author} {\bibinfo {author} {\bibfnamefont {A.}~\bibnamefont
  {Dymarsky}},\ }\href@noop {} {\enquote {\bibinfo {title} {Bound on eigenstate
  thermalization from transport},}\ }\ \bibinfo {note}
  {arXiv:1804.08626}\BibitemShut {NoStop}%
\bibitem [{\citenamefont {Reimann}(2018{\natexlab{a}})}]{Reimann2018a}%
  \BibitemOpen
  \bibfield  {author} {\bibinfo {author} {\bibfnamefont {P.}~\bibnamefont
  {Reimann}},\ }\bibfield  {title} {\enquote {\bibinfo {title} {Dynamical
  typicality approach to eigenstate thermalization},}\ }\href {\doibase
  10.1103/PhysRevLett.120.230601} {\bibfield  {journal} {\bibinfo  {journal}
  {Phys. Rev. Lett.}\ }\textbf {\bibinfo {volume} {120}},\ \bibinfo {pages}
  {230601} (\bibinfo {year} {2018}{\natexlab{a}})}\BibitemShut {NoStop}%
\bibitem [{\citenamefont {Reimann}(2018{\natexlab{b}})}]{Reimann2018b}%
  \BibitemOpen
  \bibfield  {author} {\bibinfo {author} {\bibfnamefont {P.}~\bibnamefont
  {Reimann}},\ }\bibfield  {title} {\enquote {\bibinfo {title} {Dynamical
  typicality of isolated many-body quantum systems},}\ }\href {\doibase
  10.1103/PhysRevE.97.062129} {\bibfield  {journal} {\bibinfo  {journal} {Phys.
  Rev. E}\ }\textbf {\bibinfo {volume} {97}},\ \bibinfo {pages} {062129}
  (\bibinfo {year} {2018}{\natexlab{b}})}\BibitemShut {NoStop}%
\bibitem [{\citenamefont {Monnai}(2013)}]{Monnai2013}%
  \BibitemOpen
  \bibfield  {author} {\bibinfo {author} {\bibfnamefont {Takaaki}\ \bibnamefont
  {Monnai}},\ }\bibfield  {title} {\enquote {\bibinfo {title} {Generic
  evaluation of relaxation time for quantum many-body systems: Analysis of the
  system size dependence},}\ }\href {\doibase 10.7566/JPSJ.82.044006}
  {\bibfield  {journal} {\bibinfo  {journal} {J. Phys. Soc. Jpn.}\ }\textbf
  {\bibinfo {volume} {82}},\ \bibinfo {pages} {044006} (\bibinfo {year}
  {2013})}\BibitemShut {NoStop}%
\bibitem [{\citenamefont {Goldstein}\ \emph {et~al.}(2013)\citenamefont
  {Goldstein}, \citenamefont {Hara},\ and\ \citenamefont
  {Tasaki}}]{Goldstein2013}%
  \BibitemOpen
  \bibfield  {author} {\bibinfo {author} {\bibfnamefont {Sheldon}\ \bibnamefont
  {Goldstein}}, \bibinfo {author} {\bibfnamefont {Takashi}\ \bibnamefont
  {Hara}}, \ and\ \bibinfo {author} {\bibfnamefont {Hal}\ \bibnamefont
  {Tasaki}},\ }\bibfield  {title} {\enquote {\bibinfo {title} {Time scales in
  the approach to equilibrium of macroscopic quantum systems},}\ }\href
  {\doibase 10.1103/PhysRevLett.111.140401} {\bibfield  {journal} {\bibinfo
  {journal} {Phys. Rev. Lett.}\ }\textbf {\bibinfo {volume} {111}},\ \bibinfo
  {pages} {140401} (\bibinfo {year} {2013})}\BibitemShut {NoStop}%
\bibitem [{\citenamefont {Malabarba}\ \emph {et~al.}(2014)\citenamefont
  {Malabarba}, \citenamefont {Garc\'{\i}a-Pintos}, \citenamefont {Linden},
  \citenamefont {Farrelly},\ and\ \citenamefont {Short}}]{Malabarba2014}%
  \BibitemOpen
  \bibfield  {author} {\bibinfo {author} {\bibfnamefont {Artur S.~L.}\
  \bibnamefont {Malabarba}}, \bibinfo {author} {\bibfnamefont {Luis~Pedro}\
  \bibnamefont {Garc\'{\i}a-Pintos}}, \bibinfo {author} {\bibfnamefont {Noah}\
  \bibnamefont {Linden}}, \bibinfo {author} {\bibfnamefont {Terence~C.}\
  \bibnamefont {Farrelly}}, \ and\ \bibinfo {author} {\bibfnamefont
  {Anthony~J.}\ \bibnamefont {Short}},\ }\bibfield  {title} {\enquote {\bibinfo
  {title} {Quantum systems equilibrate rapidly for most observables},}\ }\href
  {\doibase 10.1103/PhysRevE.90.012121} {\bibfield  {journal} {\bibinfo
  {journal} {Phys. Rev. E}\ }\textbf {\bibinfo {volume} {90}},\ \bibinfo
  {pages} {012121} (\bibinfo {year} {2014})}\BibitemShut {NoStop}%
\bibitem [{\citenamefont {Goldstein}\ \emph {et~al.}(2015)\citenamefont
  {Goldstein}, \citenamefont {Hara},\ and\ \citenamefont
  {Tasaki}}]{Goldstein2015}%
  \BibitemOpen
  \bibfield  {author} {\bibinfo {author} {\bibfnamefont {Sheldon}\ \bibnamefont
  {Goldstein}}, \bibinfo {author} {\bibfnamefont {Takashi}\ \bibnamefont
  {Hara}}, \ and\ \bibinfo {author} {\bibfnamefont {Hal}\ \bibnamefont
  {Tasaki}},\ }\bibfield  {title} {\enquote {\bibinfo {title} {Extremely quick
  thermalization in a macroscopic quantum system for a typical nonequilibrium
  subspace},}\ }\href {http://stacks.iop.org/1367-2630/17/i=4/a=045002}
  {\bibfield  {journal} {\bibinfo  {journal} {New J. Phys.}\ }\textbf {\bibinfo
  {volume} {17}},\ \bibinfo {pages} {045002} (\bibinfo {year}
  {2015})}\BibitemShut {NoStop}%
\bibitem [{\citenamefont {Reimann}(2016)}]{Reimann2016}%
  \BibitemOpen
  \bibfield  {author} {\bibinfo {author} {\bibfnamefont {Peter}\ \bibnamefont
  {Reimann}},\ }\bibfield  {title} {\enquote {\bibinfo {title} {Typical fast
  thermalization processes in closed many-body systems},}\ }\href {\doibase
  http://dx.doi.org/10.1038/ncomms10821} {\bibfield  {journal} {\bibinfo
  {journal} {Nat. Comm.}\ }\textbf {\bibinfo {volume} {7}},\ \bibinfo {pages}
  {10821} (\bibinfo {year} {2016})}\BibitemShut {NoStop}%
\bibitem [{\citenamefont {Garc\'{\i}a-Pintos}\ \emph
  {et~al.}(2017)\citenamefont {Garc\'{\i}a-Pintos}, \citenamefont {Linden},
  \citenamefont {Malabarba}, \citenamefont {Short},\ and\ \citenamefont
  {Winter}}]{Pintos2017}%
  \BibitemOpen
  \bibfield  {author} {\bibinfo {author} {\bibfnamefont {Luis~Pedro}\
  \bibnamefont {Garc\'{\i}a-Pintos}}, \bibinfo {author} {\bibfnamefont {Noah}\
  \bibnamefont {Linden}}, \bibinfo {author} {\bibfnamefont {Artur S.~L.}\
  \bibnamefont {Malabarba}}, \bibinfo {author} {\bibfnamefont {Anthony~J.}\
  \bibnamefont {Short}}, \ and\ \bibinfo {author} {\bibfnamefont {Andreas}\
  \bibnamefont {Winter}},\ }\bibfield  {title} {\enquote {\bibinfo {title}
  {Equilibration time scales of physically relevant observables},}\ }\href
  {\doibase 10.1103/PhysRevX.7.031027} {\bibfield  {journal} {\bibinfo
  {journal} {Phys. Rev. X}\ }\textbf {\bibinfo {volume} {7}},\ \bibinfo {pages}
  {031027} (\bibinfo {year} {2017})}\BibitemShut {NoStop}%
\bibitem [{\citenamefont {de~Oliveira}\ \emph {et~al.}(2018)\citenamefont
  {de~Oliveira}, \citenamefont {Charalambous}, \citenamefont {Jonathan},
  \citenamefont {Lewenstein},\ and\ \citenamefont {Riera}}]{Oliveira2018}%
  \BibitemOpen
  \bibfield  {author} {\bibinfo {author} {\bibfnamefont {Thiago~R}\
  \bibnamefont {de~Oliveira}}, \bibinfo {author} {\bibfnamefont {Christos}\
  \bibnamefont {Charalambous}}, \bibinfo {author} {\bibfnamefont {Daniel}\
  \bibnamefont {Jonathan}}, \bibinfo {author} {\bibfnamefont {Maciej}\
  \bibnamefont {Lewenstein}}, \ and\ \bibinfo {author} {\bibfnamefont {Arnau}\
  \bibnamefont {Riera}},\ }\bibfield  {title} {\enquote {\bibinfo {title}
  {Equilibration time scales in closed many-body quantum systems},}\ }\href
  {http://stacks.iop.org/1367-2630/20/i=3/a=033032} {\bibfield  {journal}
  {\bibinfo  {journal} {New J. Phys.}\ }\textbf {\bibinfo {volume} {20}},\
  \bibinfo {pages} {033032} (\bibinfo {year} {2018})}\BibitemShut {NoStop}%
\bibitem [{\citenamefont {Dymarsky}({\natexlab{b}})}]{DymarskyARXIVThouless}%
  \BibitemOpen
  \bibfield  {author} {\bibinfo {author} {\bibfnamefont {A.}~\bibnamefont
  {Dymarsky}},\ }\href@noop {} {\enquote {\bibinfo {title} {Mechanism of slow
  equilibration of isolated quantum systems},}\ } \ \bibinfo
  {note} {arXiv:1806.04187}\BibitemShut {NoStop}%
\bibitem [{\citenamefont {Chan}\ \emph {et~al.}(2018)\citenamefont {Chan},
  \citenamefont {De~Luca},\ and\ \citenamefont {Chalker}}]{Chan2018}%
  \BibitemOpen
  \bibfield  {author} {\bibinfo {author} {\bibfnamefont {A.}~\bibnamefont
  {Chan}}, \bibinfo {author} {\bibfnamefont {A.}~\bibnamefont {De~Luca}}, \
  and\ \bibinfo {author} {\bibfnamefont {J.~T.}\ \bibnamefont {Chalker}},\
  }\bibfield  {title} {\enquote {\bibinfo {title} {Spectral statistics in
  spatially extended chaotic quantum many-body systems},}\ }\href {\doibase
  10.1103/PhysRevLett.121.060601} {\bibfield  {journal} {\bibinfo  {journal}
  {Phys. Rev. Lett.}\ }\textbf {\bibinfo {volume} {121}},\ \bibinfo {pages}
  {060601} (\bibinfo {year} {2018})}\BibitemShut {NoStop}%
\bibitem [{\citenamefont {Bertini}\ \emph {et~al.}(2018)\citenamefont
  {Bertini}, \citenamefont {Kos},\ and\ \citenamefont {Prosen}}]{Bertini2018}%
  \BibitemOpen
  \bibfield  {author} {\bibinfo {author} {\bibfnamefont {Bruno}\ \bibnamefont
  {Bertini}}, \bibinfo {author} {\bibfnamefont {Pavel}\ \bibnamefont {Kos}}, \
  and\ \bibinfo {author} {\bibfnamefont {Toma{\v z}}\ \bibnamefont {Prosen}},\ }\bibfield  {title}
  {\enquote {\bibinfo {title} {Exact spectral form factor in a minimal model of
  many-body quantum chaos},}\ }\href {\doibase 10.1103/PhysRevLett.121.264101}
  {\bibfield  {journal} {\bibinfo  {journal} {Phys. Rev. Lett.}\ }\textbf
  {\bibinfo {volume} {121}},\ \bibinfo {pages} {264101} (\bibinfo {year}
  {2018})}\BibitemShut {NoStop}%
\bibitem [{\citenamefont {Schiulaz}\ \emph {et~al.}()\citenamefont {Schiulaz},
  \citenamefont {Torres-Herrera},\ and\ \citenamefont
  {Santos}}]{SchiulazARXIV}%
  \BibitemOpen
  \bibfield  {author} {\bibinfo {author} {\bibfnamefont {Mauro}\ \bibnamefont
  {Schiulaz}}, \bibinfo {author} {\bibfnamefont {E.~J.}\ \bibnamefont
  {Torres-Herrera}}, \ and\ \bibinfo {author} {\bibfnamefont {Lea~F.}\
  \bibnamefont {Santos}},\ }\href@noop {} {\enquote {\bibinfo {title} {Thouless
  and relaxation timescales in many-body quantum systems},}\ }\bibinfo {note}
  {arXiv:1807.07577}\BibitemShut {NoStop}%
\bibitem [{\citenamefont {Borgonovi}\ \emph {et~al.}(2019)\citenamefont
  {Borgonovi}, \citenamefont {Izrailev},\ and\ \citenamefont
  {Santos}}]{Borgonovi2019}%
  \BibitemOpen
  \bibfield  {author} {\bibinfo {author} {\bibfnamefont {Fausto}\ \bibnamefont
  {Borgonovi}}, \bibinfo {author} {\bibfnamefont {Felix~M.}\ \bibnamefont
  {Izrailev}}, \ and\ \bibinfo {author} {\bibfnamefont {Lea~F.}\ \bibnamefont
  {Santos}},\ }\bibfield  {title} {\enquote {\bibinfo {title} {Exponentially
  fast dynamics of chaotic many-body systems},}\ }\href {\doibase
  10.1103/PhysRevE.99.010101} {\bibfield  {journal} {\bibinfo  {journal} {Phys.
  Rev. E}\ }\textbf {\bibinfo {volume} {99}},\ \bibinfo {pages} {010101}
  (\bibinfo {year} {2019})}\BibitemShut {NoStop}%
\bibitem [{\citenamefont {Berry}(1983)}]{BerryProceed1981}%
  \BibitemOpen
  \bibfield  {author} {\bibinfo {author} {\bibfnamefont {M.~V.}\ \bibnamefont
  {Berry}},\ }\bibfield  {title} {\enquote {\bibinfo {title} {Semi-classical
  mechanics of regular and irregular motion},}\ }in\ \href@noop {} {\emph
  {\bibinfo {booktitle} {Les Houches Summer School 1981 on Chaotic Behaviour of
  Deterministic Systems}}},\ \bibinfo {editor} {edited by\ \bibinfo {editor}
  {\bibfnamefont {G.}~\bibnamefont {Iooss}}, \bibinfo {editor} {\bibfnamefont
  {H.G.}\ \bibnamefont {Helleman}}, \ and\ \bibinfo {editor} {\bibfnamefont
  {R.}~\bibnamefont {Stora}}}\ (\bibinfo  {publisher} {North-Holland},\
  \bibinfo {address} {Amsterdam},\ \bibinfo {year} {1983})\ p.\ \bibinfo
  {pages} {171}\BibitemShut {NoStop}%
\bibitem{footSurv}% [{foo({\natexlab{a}})}]{footSurv}%
 % \BibitemOpen
  %\href@noop {} {} ({\natexlab{a}}),\ \bibinfo {note} 
  {The survival probability
  is also called return probability, autocorrelation function, or fidelity. It
  has also been called ``Loschmidt echo'', but this term is not appropriate
  since the survival probability does not involve any time
  reversal~\cite{Goussev2012}.}%\BibitemShut {Stop}%
\bibitem{footNote} %[{foo({\natexlab{b}})}]{footNote}%
%  \BibitemOpen
 % \href@noop {} {} ({\natexlab{b}}),\ \bibinfo {note} 
 {A hole appears for
  observables that decrease in time and a bulge in the case of quantities, such
  as entropies, that increase in time~\cite{Torres2017}.}%\BibitemShut {Stop}%
\bibitem [{\citenamefont {Leviandier}\ \emph {et~al.}(1986)\citenamefont
  {Leviandier}, \citenamefont {Lombardi}, \citenamefont {Jost},\ and\
  \citenamefont {Pique}}]{Leviandier1986}%
  \BibitemOpen
  \bibfield  {author} {\bibinfo {author} {\bibfnamefont {Luc}\ \bibnamefont
  {Leviandier}}, \bibinfo {author} {\bibfnamefont {Maurice}\ \bibnamefont
  {Lombardi}}, \bibinfo {author} {\bibfnamefont {R\'emi}\ \bibnamefont {Jost}},
  \ and\ \bibinfo {author} {\bibfnamefont {Jean~Paul}\ \bibnamefont {Pique}},\
  }\bibfield  {title} {\enquote {\bibinfo {title} {Fourier transform: A tool to
  measure statistical level properties in very complex spectra},}\ }\href
  {\doibase 10.1103/PhysRevLett.56.2449} {\bibfield  {journal} {\bibinfo
  {journal} {Phys. Rev. Lett.}\ }\textbf {\bibinfo {volume} {56}},\ \bibinfo
  {pages} {2449--2452} (\bibinfo {year} {1986})}\BibitemShut {NoStop}%
\bibitem [{\citenamefont {Guhr}\ and\ \citenamefont
  {Weidenm\"uller}(1990)}]{Guhr1990}%
  \BibitemOpen
  \bibfield  {author} {\bibinfo {author} {\bibfnamefont {T.}~\bibnamefont
  {Guhr}}\ and\ \bibinfo {author} {\bibfnamefont {H.A.}\ \bibnamefont
  {Weidenm\"uller}},\ }\bibfield  {title} {\enquote {\bibinfo {title}
  {Correlations in anticrossing spectra and scattering theory. analytical
  aspects},}\ }\href {\doibase http://dx.doi.org/10.1016/0301-0104(90)90003-R}
  {\bibfield  {journal} {\bibinfo  {journal} {Chem. Phys.}\ }\textbf {\bibinfo
  {volume} {146}},\ \bibinfo {pages} {21 -- 38} (\bibinfo {year}
  {1990})}\BibitemShut {NoStop}%
\bibitem [{\citenamefont {Wilkie}\ and\ \citenamefont
  {Brumer}(1991)}]{Wilkie1991}%
  \BibitemOpen
  \bibfield  {author} {\bibinfo {author} {\bibfnamefont {Joshua}\ \bibnamefont
  {Wilkie}}\ and\ \bibinfo {author} {\bibfnamefont {Paul}\ \bibnamefont
  {Brumer}},\ }\bibfield  {title} {\enquote {\bibinfo {title} {Time-dependent
  manifestations of quantum chaos},}\ }\href {\doibase
  10.1103/PhysRevLett.67.1185} {\bibfield  {journal} {\bibinfo  {journal}
  {Phys. Rev. Lett.}\ }\textbf {\bibinfo {volume} {67}},\ \bibinfo {pages}
  {1185--1188} (\bibinfo {year} {1991})}\BibitemShut {NoStop}%
\bibitem [{\citenamefont {Alhassid}\ and\ \citenamefont
  {Levine}(1992)}]{Alhassid1992}%
  \BibitemOpen
  \bibfield  {author} {\bibinfo {author} {\bibfnamefont {Y.}~\bibnamefont
  {Alhassid}}\ and\ \bibinfo {author} {\bibfnamefont {R.~D.}\ \bibnamefont
  {Levine}},\ }\bibfield  {title} {\enquote {\bibinfo {title} {Spectral
  autocorrelation function in the statistical theory of energy levels},}\
  }\href {\doibase 10.1103/PhysRevA.46.4650} {\bibfield  {journal} {\bibinfo
  {journal} {Phys. Rev. A}\ }\textbf {\bibinfo {volume} {46}},\ \bibinfo
  {pages} {4650--4653} (\bibinfo {year} {1992})}\BibitemShut {NoStop}%
\bibitem [{\citenamefont {Torres-Herrera}\ and\ \citenamefont
  {Santos}(2017{\natexlab{a}})}]{Torres2017}%
  \BibitemOpen
  \bibfield  {author} {\bibinfo {author} {\bibfnamefont {E.~J.}\ \bibnamefont
  {Torres-Herrera}}\ and\ \bibinfo {author} {\bibfnamefont {Lea~F.}\
  \bibnamefont {Santos}},\ }\bibfield  {title} {\enquote {\bibinfo {title}
  {Extended nonergodic states in disordered many-body quantum systems},}\
  }\href {\doibase 10.1002/andp.201600284} {\bibfield  {journal} {\bibinfo
  {journal} {Ann. Phys. (Berlin)}\ }\textbf {\bibinfo {volume} {529}},\
  \bibinfo {pages} {1600284} (\bibinfo {year}
  {2017}{\natexlab{a}})}\BibitemShut {NoStop}%
\bibitem [{\citenamefont {Torres-Herrera}\ and\ \citenamefont
  {Santos}(2017{\natexlab{b}})}]{Torres2017Philo}%
  \BibitemOpen
  \bibfield  {author} {\bibinfo {author} {\bibfnamefont {E.~J.}\ \bibnamefont
  {Torres-Herrera}}\ and\ \bibinfo {author} {\bibfnamefont {Lea~F.}\
  \bibnamefont {Santos}},\ }\bibfield  {title} {\enquote {\bibinfo {title}
  {Dynamical manifestations of quantum chaos: correlation hole and bulge},}\
  }\href {\doibase 10.1098/rsta.2016.0434} {\bibfield  {journal} {\bibinfo
  {journal} {Philos. Trans. Royal Soc. A}\ }\textbf {\bibinfo {volume} {375}},\
  \bibinfo {pages} {20160434} (\bibinfo {year}
  {2017}{\natexlab{b}})}\BibitemShut {NoStop}%
\bibitem [{\citenamefont {Torres-Herrera}\ \emph {et~al.}(2018)\citenamefont
  {Torres-Herrera}, \citenamefont {Garc\'{\i}a-Garc\'{\i}a},\ and\
  \citenamefont {Santos}}]{Torres2018}%
  \BibitemOpen
  \bibfield  {author} {\bibinfo {author} {\bibfnamefont {E.~J.}\ \bibnamefont
  {Torres-Herrera}}, \bibinfo {author} {\bibfnamefont {Antonio~M.}\
  \bibnamefont {Garc\'{\i}a-Garc\'{\i}a}}, \ and\ \bibinfo {author}
  {\bibfnamefont {Lea~F.}\ \bibnamefont {Santos}},\ }\bibfield  {title}
  {\enquote {\bibinfo {title} {Generic dynamical features of quenched
  interacting quantum systems: Survival probability, density imbalance, and
  out-of-time-ordered correlator},}\ }\href {\doibase
  10.1103/PhysRevB.97.060303} {\bibfield  {journal} {\bibinfo  {journal} {Phys.
  Rev. B}\ }\textbf {\bibinfo {volume} {97}},\ \bibinfo {pages} {060303}
  (\bibinfo {year} {2018})}\BibitemShut {NoStop}%
\bibitem [{\citenamefont {Torres-Herrera}\ and\ \citenamefont
  {Santos}(2019)}]{Torres2019}%
  \BibitemOpen
  \bibfield  {author} {\bibinfo {author} {\bibfnamefont {E.~J.}\ \bibnamefont
  {Torres-Herrera}}\ and\ \bibinfo {author} {\bibfnamefont {Lea~F.}\
  \bibnamefont {Santos}},\ }\bibfield  {title} {\enquote {\bibinfo {title}
  {Signatures of chaos and thermalization in the dynamics of many-body quantum
  systems},}\ }\href@noop {} {\bibfield  {journal} {\bibinfo  {journal} {Eur.
  Phys. J. Spec. Top.}\ }\textbf {\bibinfo {volume} {227}},\ \bibinfo {pages}
  {1897} (\bibinfo {year} {2019})}\BibitemShut {NoStop}%
\bibitem [{\citenamefont {Cotler}\ \emph {et~al.}(2017)\citenamefont {Cotler},
  \citenamefont {Hunter-Jones}, \citenamefont {Liu},\ and\ \citenamefont
  {Yoshida}}]{Cotler2017GUE}%
  \BibitemOpen
  \bibfield  {author} {\bibinfo {author} {\bibfnamefont {J.}~\bibnamefont
  {Cotler}}, \bibinfo {author} {\bibfnamefont {N.}~\bibnamefont
  {Hunter-Jones}}, \bibinfo {author} {\bibfnamefont {J.}~\bibnamefont {Liu}}, \
  and\ \bibinfo {author} {\bibfnamefont {B.}~\bibnamefont {Yoshida}},\
  }\bibfield  {title} {\enquote {\bibinfo {title} {Chaos, complexity, and
  random matrices},}\ }\href {\doibase 10.1007/JHEP11(2017)048} {\bibfield
  {journal} {\bibinfo  {journal} {J. High Energy Phys.}\ }\textbf {\bibinfo
  {volume} {2017}},\ \bibinfo {pages} {48} (\bibinfo {year}
  {2017})}\BibitemShut {NoStop}%
\bibitem [{\citenamefont {Gharibyan}\ \emph {et~al.}(2018)\citenamefont
  {Gharibyan}, \citenamefont {Hanada}, \citenamefont {Shenker},\ and\
  \citenamefont {Tezuka}}]{Gharibyan2018}%
  \BibitemOpen
  \bibfield  {author} {\bibinfo {author} {\bibfnamefont {Hrant}\ \bibnamefont
  {Gharibyan}}, \bibinfo {author} {\bibfnamefont {Masanori}\ \bibnamefont
  {Hanada}}, \bibinfo {author} {\bibfnamefont {Stephen~H.}\ \bibnamefont
  {Shenker}}, \ and\ \bibinfo {author} {\bibfnamefont {Masaki}\ \bibnamefont
  {Tezuka}},\ }\bibfield  {title} {\enquote {\bibinfo {title} {Onset of random
  matrix behavior in scrambling systems},}\ }\href {\doibase
  10.1007/JHEP07(2018)124} {\bibfield  {journal} {\bibinfo  {journal} {Journal
  of High Energy Physics}\ }\textbf {\bibinfo {volume} {2018}},\ \bibinfo
  {pages} {124} (\bibinfo {year} {2018})}\BibitemShut {NoStop}%
\bibitem [{\citenamefont {Nosaka}\ \emph {et~al.}(2018)\citenamefont {Nosaka},
  \citenamefont {Rosa},\ and\ \citenamefont {Yoon}}]{Nosaka2018}%
  \BibitemOpen
  \bibfield  {author} {\bibinfo {author} {\bibfnamefont {Tomoki}\ \bibnamefont
  {Nosaka}}, \bibinfo {author} {\bibfnamefont {Dario}\ \bibnamefont {Rosa}}, \
  and\ \bibinfo {author} {\bibfnamefont {Junggi}\ \bibnamefont {Yoon}},\
  }\bibfield  {title} {\enquote {\bibinfo {title} {The {T}houless time for
  mass-deformed {SYK}},}\ }\href {\doibase 10.1007/JHEP09(2018)041} {\bibfield
  {journal} {\bibinfo  {journal} {Journal of High Energy Physics}\ }\textbf
  {\bibinfo {volume} {2018}},\ \bibinfo {pages} {41} (\bibinfo {year}
  {2018})}\BibitemShut {NoStop}%
\bibitem [{\citenamefont {Brody}\ \emph {et~al.}(1981)\citenamefont {Brody},
  \citenamefont {Flores}, \citenamefont {French}, \citenamefont {Mello},
  \citenamefont {Pandey},\ and\ \citenamefont {Wong}}]{Brody1981}%
  \BibitemOpen
  \bibfield  {author} {\bibinfo {author} {\bibfnamefont {T.~A.}\ \bibnamefont
  {Brody}}, \bibinfo {author} {\bibfnamefont {J.}~\bibnamefont {Flores}},
  \bibinfo {author} {\bibfnamefont {J.~B.}\ \bibnamefont {French}}, \bibinfo
  {author} {\bibfnamefont {P.~A.}\ \bibnamefont {Mello}}, \bibinfo {author}
  {\bibfnamefont {A.}~\bibnamefont {Pandey}}, \ and\ \bibinfo {author}
  {\bibfnamefont {S.~S.~M.}\ \bibnamefont {Wong}},\ }\bibfield  {title}
  {\enquote {\bibinfo {title} {Random-matrix physics: spectrum and strength
  fluctuations},}\ }\href {\doibase 10.1103/RevModPhys.53.385} {\bibfield
  {journal} {\bibinfo  {journal} {Rev. Mod. Phys.}\ }\textbf {\bibinfo {volume}
  {53}},\ \bibinfo {pages} {385} (\bibinfo {year} {1981})}\BibitemShut
  {NoStop}%
\bibitem{TorresARXIV} E. J. Torres-Herrera, J. A. M\'endez-Berm\'udez, and Lea F. Santos, ``Level Repulsion and Dynamics in the Finite One-Dimensional Anderson Model'', arXiv: 1904.11989.
\bibitem [{\citenamefont {Dicke}(1954)}]{Dicke1954}%
  \BibitemOpen
  \bibfield  {author} {\bibinfo {author} {\bibfnamefont {R.~H.}\ \bibnamefont
  {Dicke}},\ }\bibfield  {title} {\enquote {\bibinfo {title} {Coherence in
  spontaneous radiation processes},}\ }\href {\doibase 10.1103/PhysRev.93.99}
  {\bibfield  {journal} {\bibinfo  {journal} {Phys. Rev.}\ }\textbf {\bibinfo
  {volume} {93}},\ \bibinfo {pages} {99--110} (\bibinfo {year}
  {1954})}\BibitemShut {NoStop}%
\bibitem [{\citenamefont {Garraway}(2011)}]{Garraway2011}%
  \BibitemOpen
  \bibfield  {author} {\bibinfo {author} {\bibfnamefont {Barry~M.}\
  \bibnamefont {Garraway}},\ }\bibfield  {title} {\enquote {\bibinfo {title}
  {The dicke model in quantum optics: Dicke model revisited},}\ }\href
  {\doibase 10.1098/rsta.2010.0333} {\bibfield  {journal} {\bibinfo  {journal}
  {Philos. Trans. Royal Soc. A}\ }\textbf {\bibinfo {volume} {369}},\ \bibinfo
  {pages} {1137} (\bibinfo {year} {2011})}\BibitemShut {NoStop}%
\bibitem [{\citenamefont {Baumann}\ \emph {et~al.}(2010)\citenamefont
  {Baumann}, \citenamefont {Guerlin}, \citenamefont {Brennecke},\ and\
  \citenamefont {Esslinger}}]{Baumann2010}%
  \BibitemOpen
  \bibfield  {author} {\bibinfo {author} {\bibfnamefont {Kristian}\
  \bibnamefont {Baumann}}, \bibinfo {author} {\bibfnamefont {Christine}\
  \bibnamefont {Guerlin}}, \bibinfo {author} {\bibfnamefont {Ferdinand}\
  \bibnamefont {Brennecke}}, \ and\ \bibinfo {author} {\bibfnamefont {Tilman}\
  \bibnamefont {Esslinger}},\ }\bibfield  {title} {\enquote {\bibinfo {title}
  {{D}icke quantum phase transition with a superfluid gas in an optical
  cavity},}\ }\href {\doibase doi:10.1038/nature09009} {\bibfield  {journal}
  {\bibinfo  {journal} {Nature (London)}\ }\textbf {\bibinfo {volume} {464}},\
  \bibinfo {pages} {1301} (\bibinfo {year} {2010})}\BibitemShut {NoStop}%
\bibitem [{\citenamefont {Baumann}\ \emph {et~al.}(2011)\citenamefont
  {Baumann}, \citenamefont {Mottl}, \citenamefont {Brennecke},\ and\
  \citenamefont {Esslinger}}]{Baumann2011}%
  \BibitemOpen
  \bibfield  {author} {\bibinfo {author} {\bibfnamefont {K.}~\bibnamefont
  {Baumann}}, \bibinfo {author} {\bibfnamefont {R.}~\bibnamefont {Mottl}},
  \bibinfo {author} {\bibfnamefont {F.}~\bibnamefont {Brennecke}}, \ and\
  \bibinfo {author} {\bibfnamefont {T.}~\bibnamefont {Esslinger}},\ }\bibfield
  {title} {\enquote {\bibinfo {title} {Exploring symmetry breaking at the
  {D}icke quantum phase transition},}\ }\href {\doibase
  10.1103/PhysRevLett.107.140402} {\bibfield  {journal} {\bibinfo  {journal}
  {Phys. Rev. Lett.}\ }\textbf {\bibinfo {volume} {107}},\ \bibinfo {pages}
  {140402} (\bibinfo {year} {2011})}\BibitemShut {NoStop}%
\bibitem [{\citenamefont {Lewenkopf}\ \emph {et~al.}(1991)\citenamefont
  {Lewenkopf}, \citenamefont {Nemes}, \citenamefont {Marvulle}, \citenamefont
  {Pato},\ and\ \citenamefont {Wreszinski}}]{Lewenkopf1991}%
  \BibitemOpen
  \bibfield  {author} {\bibinfo {author} {\bibfnamefont {C.H}\ \bibnamefont
  {Lewenkopf}}, \bibinfo {author} {\bibfnamefont {M.C}\ \bibnamefont {Nemes}},
  \bibinfo {author} {\bibfnamefont {V}~\bibnamefont {Marvulle}}, \bibinfo
  {author} {\bibfnamefont {M.P}\ \bibnamefont {Pato}}, \ and\ \bibinfo {author}
  {\bibfnamefont {W.F}\ \bibnamefont {Wreszinski}},\ }\bibfield  {title}
  {\enquote {\bibinfo {title} {Level statistics transitions in the spin-boson
  model},}\ }\href {\doibase https://doi.org/10.1016/0375-9601(91)90575-S}
  {\bibfield  {journal} {\bibinfo  {journal} {Phys. Lett. A}\ }\textbf
  {\bibinfo {volume} {155}},\ \bibinfo {pages} {113 -- 116} (\bibinfo {year}
  {1991})}\BibitemShut {NoStop}%
\bibitem [{\citenamefont {Emary}\ and\ \citenamefont
  {Brandes}(2003{\natexlab{a}})}]{Emary2003PRL}%
  \BibitemOpen
  \bibfield  {author} {\bibinfo {author} {\bibfnamefont {Clive}\ \bibnamefont
  {Emary}}\ and\ \bibinfo {author} {\bibfnamefont {Tobias}\ \bibnamefont
  {Brandes}},\ }\bibfield  {title} {\enquote {\bibinfo {title} {Quantum chaos
  triggered by precursors of a quantum phase transition: The {D}icke model},}\
  }\href {\doibase 10.1103/PhysRevLett.90.044101} {\bibfield  {journal}
  {\bibinfo  {journal} {Phys. Rev. Lett.}\ }\textbf {\bibinfo {volume} {90}},\
  \bibinfo {pages} {044101} (\bibinfo {year} {2003}{\natexlab{a}})}\BibitemShut
  {NoStop}%
\bibitem [{\citenamefont {Emary}\ and\ \citenamefont
  {Brandes}(2003{\natexlab{b}})}]{Emary2003}%
  \BibitemOpen
  \bibfield  {author} {\bibinfo {author} {\bibfnamefont {Clive}\ \bibnamefont
  {Emary}}\ and\ \bibinfo {author} {\bibfnamefont {Tobias}\ \bibnamefont
  {Brandes}},\ }\bibfield  {title} {\enquote {\bibinfo {title} {Chaos and the
  quantum phase transition in the {D}icke model},}\ }\href {\doibase
  10.1103/PhysRevE.67.066203} {\bibfield  {journal} {\bibinfo  {journal} {Phys.
  Rev. E}\ }\textbf {\bibinfo {volume} {67}},\ \bibinfo {pages} {066203}
  (\bibinfo {year} {2003}{\natexlab{b}})}\BibitemShut {NoStop}%
\bibitem [{\citenamefont {Bastarrachea-Magnani}\ \emph
  {et~al.}(2014{\natexlab{a}})\citenamefont {Bastarrachea-Magnani},
  \citenamefont {Lerma-Hern\'andez},\ and\ \citenamefont
  {Hirsch}}]{Bastarrachea2014b}%
  \BibitemOpen
  \bibfield  {author} {\bibinfo {author} {\bibfnamefont {M.~A.}\ \bibnamefont
  {Bastarrachea-Magnani}}, \bibinfo {author} {\bibfnamefont {S.}~\bibnamefont
  {Lerma-Hern\'andez}}, \ and\ \bibinfo {author} {\bibfnamefont {J.~G.}\
  \bibnamefont {Hirsch}},\ }\bibfield  {title} {\enquote {\bibinfo {title}
  {Comparative quantum and semiclassical analysis of atom-field systems. ii.
  {C}haos and regularity},}\ }\href {\doibase 10.1103/PhysRevA.89.032102}
  {\bibfield  {journal} {\bibinfo  {journal} {Phys. Rev. A}\ }\textbf {\bibinfo
  {volume} {89}},\ \bibinfo {pages} {032102} (\bibinfo {year}
  {2014}{\natexlab{a}})}\BibitemShut {NoStop}%
\bibitem [{\citenamefont {Bastarrachea-Magnani}\ \emph
  {et~al.}(2015)\citenamefont {Bastarrachea-Magnani}, \citenamefont {del
  Carpio}, \citenamefont {Lerma-Hern\'andez},\ and\ \citenamefont
  {Hirsch}}]{Bastarrachea2015}%
  \BibitemOpen
  \bibfield  {author} {\bibinfo {author} {\bibfnamefont {Miguel~Angel}\
  \bibnamefont {Bastarrachea-Magnani}}, \bibinfo {author} {\bibfnamefont
  {Baldemar~L\'opez}\ \bibnamefont {del Carpio}}, \bibinfo {author}
  {\bibfnamefont {Sergio}\ \bibnamefont {Lerma-Hern\'andez}}, \ and\ \bibinfo
  {author} {\bibfnamefont {Jorge~G}\ \bibnamefont {Hirsch}},\ }\bibfield
  {title} {\enquote {\bibinfo {title} {Chaos in the {D}icke model: quantum and
  semiclassical analysis},}\ }\href
  {http://stacks.iop.org/1402-4896/90/i=6/a=068015} {\bibfield  {journal}
  {\bibinfo  {journal} {Phys. Scripta}\ }\textbf {\bibinfo {volume} {90}},\
  \bibinfo {pages} {068015} (\bibinfo {year} {2015})}\BibitemShut {NoStop}%
\bibitem [{\citenamefont {Bastarrachea-Magnani}\ \emph
  {et~al.}(2016)\citenamefont {Bastarrachea-Magnani}, \citenamefont
  {L\'opez-del{-}Carpio}, \citenamefont {Ch\'avez-Carlos}, \citenamefont
  {Lerma-Hern\'andez},\ and\ \citenamefont {Hirsch}}]{Bastarrachea2016PRE}%
  \BibitemOpen
  \bibfield  {author} {\bibinfo {author} {\bibfnamefont {M.~A.}\ \bibnamefont
  {Bastarrachea-Magnani}}, \bibinfo {author} {\bibfnamefont {B.}~\bibnamefont
  {L\'opez-del{-}Carpio}}, \bibinfo {author} {\bibfnamefont {J.}~\bibnamefont
  {Ch\'avez-Carlos}}, \bibinfo {author} {\bibfnamefont {S.}~\bibnamefont
  {Lerma-Hern\'andez}}, \ and\ \bibinfo {author} {\bibfnamefont {J.~G.}\
  \bibnamefont {Hirsch}},\ }\bibfield  {title} {\enquote {\bibinfo {title}
  {Delocalization and quantum chaos in atom-field systems},}\ }\href {\doibase
  10.1103/PhysRevE.93.022215} {\bibfield  {journal} {\bibinfo  {journal} {Phys.
  Rev. E}\ }\textbf {\bibinfo {volume} {93}},\ \bibinfo {pages} {022215}
  (\bibinfo {year} {2016})}\BibitemShut {NoStop}%
\bibitem [{\citenamefont {Ch\'avez-Carlos}\ \emph {et~al.}(2016)\citenamefont
  {Ch\'avez-Carlos}, \citenamefont {Bastarrachea-Magnani}, \citenamefont
  {Lerma-Hern\'andez},\ and\ \citenamefont {Hirsch}}]{Chavez2016}%
  \BibitemOpen
  \bibfield  {author} {\bibinfo {author} {\bibfnamefont {J.}~\bibnamefont
  {Ch\'avez-Carlos}}, \bibinfo {author} {\bibfnamefont {M.~A.}\ \bibnamefont
  {Bastarrachea-Magnani}}, \bibinfo {author} {\bibfnamefont {S.}~\bibnamefont
  {Lerma-Hern\'andez}}, \ and\ \bibinfo {author} {\bibfnamefont {J.~G.}\
  \bibnamefont {Hirsch}},\ }\bibfield  {title} {\enquote {\bibinfo {title}
  {Classical chaos in atom-field systems},}\ }\href {\doibase
  10.1103/PhysRevE.94.022209} {\bibfield  {journal} {\bibinfo  {journal} {Phys.
  Rev. E}\ }\textbf {\bibinfo {volume} {94}},\ \bibinfo {pages} {022209}
  (\bibinfo {year} {2016})}\BibitemShut {NoStop}%
\bibitem [{\citenamefont {Andolina}\ \emph {et~al.}()\citenamefont {Andolina},
  \citenamefont {Keck}, \citenamefont {Mari}, \citenamefont {Giovannetti},\
  and\ \citenamefont {Polini}}]{AndolinaARXIV}%
  \BibitemOpen
  \bibfield  {author} {\bibinfo {author} {\bibfnamefont {Gian~Marcello}\
  \bibnamefont {Andolina}}, \bibinfo {author} {\bibfnamefont {Maximilian}\
  \bibnamefont {Keck}}, \bibinfo {author} {\bibfnamefont {Andrea}\ \bibnamefont
  {Mari}}, \bibinfo {author} {\bibfnamefont {Vittorio}\ \bibnamefont
  {Giovannetti}}, \ and\ \bibinfo {author} {\bibfnamefont {Marco}\ \bibnamefont
  {Polini}},\ }\href@noop {} {\enquote {\bibinfo {title} {Quantum versus
  classical many-body batteries},}\ }\bibinfo {note}
  {arXiv:1812.04669}\BibitemShut {NoStop}%
\bibitem [{\citenamefont {P\'erez-Fern\'andez}\ \emph
  {et~al.}(2011{\natexlab{a}})\citenamefont {P\'erez-Fern\'andez},
  \citenamefont {Rela\~no}, \citenamefont {Arias}, \citenamefont {Cejnar},
  \citenamefont {Dukelsky},\ and\ \citenamefont
  {Garc\'{i}a-Ramos}}]{Fernandez2011b}%
  \BibitemOpen
  \bibfield  {author} {\bibinfo {author} {\bibfnamefont {P.}~\bibnamefont
  {P\'erez-Fern\'andez}}, \bibinfo {author} {\bibfnamefont {A.}~\bibnamefont
  {Rela\~no}}, \bibinfo {author} {\bibfnamefont {J.~M.}\ \bibnamefont {Arias}},
  \bibinfo {author} {\bibfnamefont {P.}~\bibnamefont {Cejnar}}, \bibinfo
  {author} {\bibfnamefont {J.}~\bibnamefont {Dukelsky}}, \ and\ \bibinfo
  {author} {\bibfnamefont {J.~E.}\ \bibnamefont {Garc\'{i}a-Ramos}},\
  }\bibfield  {title} {\enquote {\bibinfo {title} {Excited-state phase
  transition and onset of chaos in quantum optical models},}\ }\href {\doibase
  10.1103/PhysRevE.83.046208} {\bibfield  {journal} {\bibinfo  {journal} {Phys.
  Rev. E}\ }\textbf {\bibinfo {volume} {83}},\ \bibinfo {pages} {046208}
  (\bibinfo {year} {2011}{\natexlab{a}})}\BibitemShut {NoStop}%
\bibitem [{\citenamefont {Brandes}(2013)}]{Brandes2013}%
  \BibitemOpen
  \bibfield  {author} {\bibinfo {author} {\bibfnamefont {Tobias}\ \bibnamefont
  {Brandes}},\ }\bibfield  {title} {\enquote {\bibinfo {title} {Excited-state
  quantum phase transitions in {D}icke superradiance models},}\ }\href
  {\doibase 10.1103/PhysRevE.88.032133} {\bibfield  {journal} {\bibinfo
  {journal} {Phys. Rev. E}\ }\textbf {\bibinfo {volume} {88}},\ \bibinfo
  {pages} {032133} (\bibinfo {year} {2013})}\BibitemShut {NoStop}%
\bibitem [{\citenamefont {Bastarrachea-Magnani}\ \emph
  {et~al.}(2014{\natexlab{b}})\citenamefont {Bastarrachea-Magnani},
  \citenamefont {Lerma-Hern\'andez},\ and\ \citenamefont
  {Hirsch}}]{Bastarrachea2014a}%
  \BibitemOpen
  \bibfield  {author} {\bibinfo {author} {\bibfnamefont {M.~A.}\ \bibnamefont
  {Bastarrachea-Magnani}}, \bibinfo {author} {\bibfnamefont {S.}~\bibnamefont
  {Lerma-Hern\'andez}}, \ and\ \bibinfo {author} {\bibfnamefont {J.~G.}\
  \bibnamefont {Hirsch}},\ }\bibfield  {title} {\enquote {\bibinfo {title}
  {Comparative quantum and semiclassical analysis of atom-field systems. {I}.
  {D}ensity of states and excited-state quantum phase transitions},}\ }\href
  {\doibase 10.1103/PhysRevA.89.032101} {\bibfield  {journal} {\bibinfo
  {journal} {Phys. Rev. A}\ }\textbf {\bibinfo {volume} {89}},\ \bibinfo
  {pages} {032101} (\bibinfo {year} {2014}{\natexlab{b}})}\BibitemShut
  {NoStop}%
\bibitem [{\citenamefont {Larson}\ and\ \citenamefont
  {Irish}(2017)}]{Larson2017}%
  \BibitemOpen
  \bibfield  {author} {\bibinfo {author} {\bibfnamefont {Jonas}\ \bibnamefont
  {Larson}}\ and\ \bibinfo {author} {\bibfnamefont {Elinor~K}\ \bibnamefont
  {Irish}},\ }\bibfield  {title} {\enquote {\bibinfo {title} {Some remarks on
  superradiant phase transitions in light-matter systems},}\ }\href
  {http://stacks.iop.org/1751-8121/50/i=17/a=174002} {\bibfield  {journal}
  {\bibinfo  {journal} {J. Phys. A}\ }\textbf {\bibinfo {volume} {50}},\
  \bibinfo {pages} {174002} (\bibinfo {year} {2017})}\BibitemShut {NoStop}%
\bibitem [{\citenamefont {P\'erez-Fern\'andez}\ \emph
  {et~al.}(2011{\natexlab{b}})\citenamefont {P\'erez-Fern\'andez},
  \citenamefont {Cejnar}, \citenamefont {Arias}, \citenamefont {Dukelsky},
  \citenamefont {Garc\'{i}a-Ramos},\ and\ \citenamefont
  {Rela\~no}}]{Fernandez2011}%
  \BibitemOpen
  \bibfield  {author} {\bibinfo {author} {\bibfnamefont {P.}~\bibnamefont
  {P\'erez-Fern\'andez}}, \bibinfo {author} {\bibfnamefont {P.}~\bibnamefont
  {Cejnar}}, \bibinfo {author} {\bibfnamefont {J.~M.}\ \bibnamefont {Arias}},
  \bibinfo {author} {\bibfnamefont {J.}~\bibnamefont {Dukelsky}}, \bibinfo
  {author} {\bibfnamefont {J.~E.}\ \bibnamefont {Garc\'{i}a-Ramos}}, \ and\
  \bibinfo {author} {\bibfnamefont {A.}~\bibnamefont {Rela\~no}},\ }\bibfield
  {title} {\enquote {\bibinfo {title} {Quantum quench influenced by an
  excited-state phase transition},}\ }\href {\doibase
  10.1103/PhysRevA.83.033802} {\bibfield  {journal} {\bibinfo  {journal} {Phys.
  Rev. A}\ }\textbf {\bibinfo {volume} {83}},\ \bibinfo {pages} {033802}
  (\bibinfo {year} {2011}{\natexlab{b}})}\BibitemShut {NoStop}%
\bibitem [{\citenamefont {Altland}\ and\ \citenamefont
  {Haake}(2012)}]{Altland2012PRL}%
  \BibitemOpen
  \bibfield  {author} {\bibinfo {author} {\bibfnamefont {Alexander}\
  \bibnamefont {Altland}}\ and\ \bibinfo {author} {\bibfnamefont {Fritz}\
  \bibnamefont {Haake}},\ }\bibfield  {title} {\enquote {\bibinfo {title}
  {Quantum chaos and effective thermalization},}\ }\href {\doibase
  10.1103/PhysRevLett.108.073601} {\bibfield  {journal} {\bibinfo  {journal}
  {Phys. Rev. Lett.}\ }\textbf {\bibinfo {volume} {108}},\ \bibinfo {pages}
  {073601} (\bibinfo {year} {2012})}\BibitemShut {NoStop}%
\bibitem [{\citenamefont {Lerma-Hern\'andez}\ \emph {et~al.}(2018)\citenamefont
  {Lerma-Hern\'andez}, \citenamefont {Ch\'avez-Carlos}, \citenamefont
  {Bastarrachea-Magnani}, \citenamefont {Santos},\ and\ \citenamefont
  {Hirsch}}]{Lerma2018}%
  \BibitemOpen
  \bibfield  {author} {\bibinfo {author} {\bibfnamefont {Sergio}\ \bibnamefont
  {Lerma-Hern\'andez}}, \bibinfo {author} {\bibfnamefont {Jorge}\ \bibnamefont
  {Ch\'avez-Carlos}}, \bibinfo {author} {\bibfnamefont {Miguel~A.}\
  \bibnamefont {Bastarrachea-Magnani}}, \bibinfo {author} {\bibfnamefont
  {Lea~F.}\ \bibnamefont {Santos}}, \ and\ \bibinfo {author} {\bibfnamefont
  {Jorge~G.}\ \bibnamefont {Hirsch}},\ }\bibfield  {title} {\enquote {\bibinfo
  {title} {Analytical description of the survival probability of coherent
  states in regular regimes},}\ }\href
  {http://stacks.iop.org/1751-8121/51/i=47/a=475302} {\bibfield  {journal}
  {\bibinfo  {journal} {J. Phys. A}\ }\textbf {\bibinfo {volume} {51}},\
  \bibinfo {pages} {475302} (\bibinfo {year} {2018})}\BibitemShut {NoStop}%
\bibitem [{\citenamefont {Kloc}\ \emph {et~al.}(2018)\citenamefont {Kloc},
  \citenamefont {Str\'ansk\'y},\ and\ \citenamefont {Cejnar}}]{Kloc2018}%
  \BibitemOpen
  \bibfield  {author} {\bibinfo {author} {\bibfnamefont {Michal}\ \bibnamefont
  {Kloc}}, \bibinfo {author} {\bibfnamefont {Pavel}\ \bibnamefont
  {Str\'ansk\'y}}, \ and\ \bibinfo {author} {\bibfnamefont {Pavel}\
  \bibnamefont {Cejnar}},\ }\bibfield  {title} {\enquote {\bibinfo {title}
  {Quantum quench dynamics in {D}icke superradiance models},}\ }\href {\doibase
  10.1103/PhysRevA.98.013836} {\bibfield  {journal} {\bibinfo  {journal} {Phys.
  Rev. A}\ }\textbf {\bibinfo {volume} {98}},\ \bibinfo {pages} {013836}
  (\bibinfo {year} {2018})}\BibitemShut {NoStop}%
\bibitem [{\citenamefont {Ch\'avez-Carlos}\ \emph {et~al.}(2019)\citenamefont
  {Ch\'avez-Carlos}, \citenamefont {L\'opez-del Carpio}, \citenamefont
  {Bastarrachea-Magnani}, \citenamefont {Str\'ansk\'y}, \citenamefont
  {Lerma-Hern\'andez}, \citenamefont {Santos},\ and\ \citenamefont
  {Hirsch}}]{Chavez2019}%
  \BibitemOpen
  \bibfield  {author} {\bibinfo {author} {\bibfnamefont {Jorge}\ \bibnamefont
  {Ch\'avez-Carlos}}, \bibinfo {author} {\bibfnamefont {B.}~\bibnamefont
  {L\'opez-del Carpio}}, \bibinfo {author} {\bibfnamefont {Miguel~A.}\
  \bibnamefont {Bastarrachea-Magnani}}, \bibinfo {author} {\bibfnamefont
  {Pavel}\ \bibnamefont {Str\'ansk\'y}}, \bibinfo {author} {\bibfnamefont
  {Sergio}\ \bibnamefont {Lerma-Hern\'andez}}, \bibinfo {author} {\bibfnamefont
  {Lea~F.}\ \bibnamefont {Santos}}, \ and\ \bibinfo {author} {\bibfnamefont
  {Jorge~G.}\ \bibnamefont {Hirsch}},\ }\bibfield  {title} {\enquote {\bibinfo
  {title} {Quantum and classical lyapunov exponents in atom-field interaction
  systems},}\ }\href {\doibase 10.1103/PhysRevLett.122.024101} {\bibfield
  {journal} {\bibinfo  {journal} {Phys. Rev. Lett.}\ }\textbf {\bibinfo
  {volume} {122}},\ \bibinfo {pages} {024101} (\bibinfo {year}
  {2019})}\BibitemShut {NoStop}%
\bibitem [{\citenamefont {Baden}\ \emph {et~al.}(2014)\citenamefont {Baden},
  \citenamefont {Arnold}, \citenamefont {Grimsmo}, \citenamefont {Parkins},\
  and\ \citenamefont {Barrett}}]{Baden2014}%
  \BibitemOpen
  \bibfield  {author} {\bibinfo {author} {\bibfnamefont {Markus~P.}\
  \bibnamefont {Baden}}, \bibinfo {author} {\bibfnamefont {Kyle~J.}\
  \bibnamefont {Arnold}}, \bibinfo {author} {\bibfnamefont {Arne~L.}\
  \bibnamefont {Grimsmo}}, \bibinfo {author} {\bibfnamefont {Scott}\
  \bibnamefont {Parkins}}, \ and\ \bibinfo {author} {\bibfnamefont {Murray~D.}\
  \bibnamefont {Barrett}},\ }\bibfield  {title} {\enquote {\bibinfo {title}
  {Realization of the {D}icke model using cavity-assisted raman transitions},}\
  }\href {\doibase 10.1103/PhysRevLett.113.020408} {\bibfield  {journal}
  {\bibinfo  {journal} {Phys. Rev. Lett.}\ }\textbf {\bibinfo {volume} {113}},\
  \bibinfo {pages} {020408} (\bibinfo {year} {2014})}\BibitemShut {NoStop}%
\bibitem [{\citenamefont {Klinder}\ \emph {et~al.}(2015)\citenamefont
  {Klinder}, \citenamefont {Ke\ss{}ler}, \citenamefont {Bakhtiari},
  \citenamefont {Thorwart},\ and\ \citenamefont {Hemmerich}}]{Klinder2015}%
  \BibitemOpen
  \bibfield  {author} {\bibinfo {author} {\bibfnamefont {J.}~\bibnamefont
  {Klinder}}, \bibinfo {author} {\bibfnamefont {H.}~\bibnamefont {Ke\ss{}ler}},
  \bibinfo {author} {\bibfnamefont {M.~Reza}\ \bibnamefont {Bakhtiari}},
  \bibinfo {author} {\bibfnamefont {M.}~\bibnamefont {Thorwart}}, \ and\
  \bibinfo {author} {\bibfnamefont {A.}~\bibnamefont {Hemmerich}},\ }\bibfield
  {title} {\enquote {\bibinfo {title} {Observation of a superradiant mott
  insulator in the {D}icke-hubbard model},}\ }\href {\doibase
  10.1103/PhysRevLett.115.230403} {\bibfield  {journal} {\bibinfo  {journal}
  {Phys. Rev. Lett.}\ }\textbf {\bibinfo {volume} {115}},\ \bibinfo {pages}
  {230403} (\bibinfo {year} {2015})}\BibitemShut {NoStop}%
\bibitem [{\citenamefont {Cohn}\ \emph {et~al.}(2018)\citenamefont {Cohn},
  \citenamefont {Safavi-Naini}, \citenamefont {Lewis-Swan}, \citenamefont
  {Bohnet}, \citenamefont {G\"arttner}, \citenamefont {Gilmore}, \citenamefont
  {Jordan}, \citenamefont {Rey}, \citenamefont {Bollinger},\ and\ \citenamefont
  {Freericks}}]{Cohn2018}%
  \BibitemOpen
  \bibfield  {author} {\bibinfo {author} {\bibfnamefont {J}~\bibnamefont
  {Cohn}}, \bibinfo {author} {\bibfnamefont {A}~\bibnamefont {Safavi-Naini}},
  \bibinfo {author} {\bibfnamefont {R~J}\ \bibnamefont {Lewis-Swan}}, \bibinfo
  {author} {\bibfnamefont {J~G}\ \bibnamefont {Bohnet}}, \bibinfo {author}
  {\bibfnamefont {M}~\bibnamefont {G\"arttner}}, \bibinfo {author}
  {\bibfnamefont {K~A}\ \bibnamefont {Gilmore}}, \bibinfo {author}
  {\bibfnamefont {J~E}\ \bibnamefont {Jordan}}, \bibinfo {author}
  {\bibfnamefont {A~M}\ \bibnamefont {Rey}}, \bibinfo {author} {\bibfnamefont
  {J~J}\ \bibnamefont {Bollinger}}, \ and\ \bibinfo {author} {\bibfnamefont
  {J~K}\ \bibnamefont {Freericks}},\ }\bibfield  {title} {\enquote {\bibinfo
  {title} {Bang-bang shortcut to adiabaticity in the dicke model as realized in
  a penning trap experiment},}\ }\href {\doibase 10.1088/1367-2630/aac3fa}
  {\bibfield  {journal} {\bibinfo  {journal} {New J. Phys.}\ }\textbf {\bibinfo
  {volume} {20}},\ \bibinfo {pages} {055013} (\bibinfo {year}
  {2018})}\BibitemShut {NoStop}%
\bibitem [{\citenamefont {Blatt}\ and\ \citenamefont {Roos}(2012)}]{Blatt2012}%
  \BibitemOpen
  \bibfield  {author} {\bibinfo {author} {\bibfnamefont {R.}~\bibnamefont
  {Blatt}}\ and\ \bibinfo {author} {\bibfnamefont {C.~F.}\ \bibnamefont
  {Roos}},\ }\bibfield  {title} {\enquote {\bibinfo {title} {Quantum
  simulations with trapped ions},}\ }\href {\doibase 10.1038/nphys2252}
  {\bibfield  {journal} {\bibinfo  {journal} {Nat. Phys.}\ }\textbf {\bibinfo
  {volume} {8}},\ \bibinfo {pages} {277--284} (\bibinfo {year}
  {2012})}\BibitemShut {NoStop}%
\bibitem [{\citenamefont {Hepp}\ and\ \citenamefont
  {Lieb}(1973{\natexlab{a}})}]{Hepp1973a}%
  \BibitemOpen
  \bibfield  {author} {\bibinfo {author} {\bibfnamefont {Klaus}\ \bibnamefont
  {Hepp}}\ and\ \bibinfo {author} {\bibfnamefont {Elliott~H}\ \bibnamefont
  {Lieb}},\ }\bibfield  {title} {\enquote {\bibinfo {title} {On the
  superradiant phase transition for molecules in a quantized radiation field:
  the {D}icke maser model},}\ }\href {\doibase
  https://doi.org/10.1016/0003-4916(73)90039-0} {\bibfield  {journal} {\bibinfo
   {journal} {Ann. Phys. (N.Y.)}\ }\textbf {\bibinfo {volume} {76}},\ \bibinfo
  {pages} {360 -- 404} (\bibinfo {year} {1973}{\natexlab{a}})}\BibitemShut
  {NoStop}%
\bibitem [{\citenamefont {Hepp}\ and\ \citenamefont
  {Lieb}(1973{\natexlab{b}})}]{Hepp1973b}%
  \BibitemOpen
  \bibfield  {author} {\bibinfo {author} {\bibfnamefont {Klaus}\ \bibnamefont
  {Hepp}}\ and\ \bibinfo {author} {\bibfnamefont {Elliott~H.}\ \bibnamefont
  {Lieb}},\ }\bibfield  {title} {\enquote {\bibinfo {title} {Equilibrium
  statistical mechanics of matter interacting with the quantized radiation
  field},}\ }\href {\doibase 10.1103/PhysRevA.8.2517} {\bibfield  {journal}
  {\bibinfo  {journal} {Phys. Rev. A}\ }\textbf {\bibinfo {volume} {8}},\
  \bibinfo {pages} {2517--2525} (\bibinfo {year}
  {1973}{\natexlab{b}})}\BibitemShut {NoStop}%
\bibitem{Casati1980} G. Casati, F. Valz-Gris, and I. Guarnieri, ``On the connection between quantization of nonintegrable systems and statistical theory of spectra'', Lett. Nuovo Cimento {\bf 28}, 279 (1980).
\bibitem{Bohigas1984}  O. Bohigas, M. J. Giannoni, and C. Schmit, ``Characterization of Chaotic Quantum Spectra and Universality of Level Fluctuation Laws'', Phys. Rev. Lett. {\bf 52}, 1 (1984).
\bibitem [{\citenamefont {Guhr}\ \emph {et~al.}(1998)\citenamefont {Guhr},
  \citenamefont {Mueller-Gr\"oeling},\ and\ \citenamefont
  {Weidenm\"uller}}]{Guhr1998}%
  \BibitemOpen
  \bibfield  {author} {\bibinfo {author} {\bibfnamefont {T.}~\bibnamefont
  {Guhr}}, \bibinfo {author} {\bibfnamefont {A.}~\bibnamefont
  {Mueller-Gr\"oeling}}, \ and\ \bibinfo {author} {\bibfnamefont {H.~A.}\
  \bibnamefont {Weidenm\"uller}},\ }\bibfield  {title} {\enquote {\bibinfo
  {title} {Random matrix theories in quantum physics: Common concepts},}\
  }\href@noop {} {\bibfield  {journal} {\bibinfo  {journal} {Phys. Rep.}\
  }\textbf {\bibinfo {volume} {299}},\ \bibinfo {pages} {189} (\bibinfo {year}
  {1998})}\BibitemShut {NoStop}%
\bibitem [{\citenamefont {Bastarrachea-Magnani}\ and\ \citenamefont
  {Hirsch}(2014)}]{Bastarrachea2014PSa}%
  \BibitemOpen
  \bibfield  {author} {\bibinfo {author} {\bibfnamefont {Miguel~A}\
  \bibnamefont {Bastarrachea-Magnani}}\ and\ \bibinfo {author} {\bibfnamefont
  {Jorge~G}\ \bibnamefont {Hirsch}},\ }\bibfield  {title} {\enquote {\bibinfo
  {title} {Efficient basis for the {D}icke model: {I}. {T}heory and convergence
  in energy},}\ }\href {http://stacks.iop.org/1402-4896/2014/i=T160/a=014005}
  {\bibfield  {journal} {\bibinfo  {journal} {Phys. Scripta}\ }\textbf
  {\bibinfo {volume} {2014}},\ \bibinfo {pages} {014005} (\bibinfo {year}
  {2014})}\BibitemShut {NoStop}%
\bibitem [{\citenamefont {Hirsch}\ and\ \citenamefont
  {Bastarrachea-Magnani}(2014)}]{Bastarrachea2014PSb}%
  \BibitemOpen
  \bibfield  {author} {\bibinfo {author} {\bibfnamefont {Jorge~G}\ \bibnamefont
  {Hirsch}}\ and\ \bibinfo {author} {\bibfnamefont {Miguel~A}\ \bibnamefont
  {Bastarrachea-Magnani}},\ }\bibfield  {title} {\enquote {\bibinfo {title}
  {Efficient basis for the {D}icke model: {II}. {W}ave function convergence and
  excited states},}\ }\href
  {http://stacks.iop.org/1402-4896/2014/i=T160/a=014018} {\bibfield  {journal}
  {\bibinfo  {journal} {Phys. Scripta}\ }\textbf {\bibinfo {volume} {2014}},\
  \bibinfo {pages} {014018} (\bibinfo {year} {2014})}\BibitemShut {NoStop}%
\bibitem [{\citenamefont {Torres-Herrera}\ and\ \citenamefont
  {Santos}(2014)}]{Torres2014PRA}%
  \BibitemOpen
  \bibfield  {author} {\bibinfo {author} {\bibfnamefont {E.~J.}\ \bibnamefont
  {Torres-Herrera}}\ and\ \bibinfo {author} {\bibfnamefont {Lea~F.}\
  \bibnamefont {Santos}},\ }\bibfield  {title} {\enquote {\bibinfo {title}
  {Quench dynamics of isolated many-body quantum systems},}\ }\href {\doibase
  10.1103/PhysRevA.89.043620} {\bibfield  {journal} {\bibinfo  {journal} {Phys.
  Rev. A}\ }\textbf {\bibinfo {volume} {89}},\ \bibinfo {pages} {043620}
  (\bibinfo {year} {2014})}\BibitemShut {NoStop}%
\bibitem [{\citenamefont {Khalfin}(1958)}]{Khalfin1958}%
  \BibitemOpen
  \bibfield  {author} {\bibinfo {author} {\bibfnamefont {L.~A.}\ \bibnamefont
  {Khalfin}},\ }\bibfield  {title} {\enquote {\bibinfo {title} {Contribution to
  the decay theory of a quasi-stationary state},}\ }\href@noop {} {\bibfield
  {journal} {\bibinfo  {journal} {Sov. Phys. JETP}\ }\textbf {\bibinfo {volume}
  {6}},\ \bibinfo {pages} {1053} (\bibinfo {year} {1958})}\BibitemShut
  {NoStop}%
\bibitem [{\citenamefont {Fonda}\ \emph {et~al.}(1978)\citenamefont {Fonda},
  \citenamefont {Ghirardi},\ and\ \citenamefont {Rimini}}]{Fonda1978}%
  \BibitemOpen
  \bibfield  {author} {\bibinfo {author} {\bibfnamefont {L.}~\bibnamefont
  {Fonda}}, \bibinfo {author} {\bibfnamefont {G.~C.}\ \bibnamefont {Ghirardi}},
  \ and\ \bibinfo {author} {\bibfnamefont {A.}~\bibnamefont {Rimini}},\
  }\bibfield  {title} {\enquote {\bibinfo {title} {Decay theory of unstable
  quantum systems},}\ }\href@noop {} {\bibfield  {journal} {\bibinfo  {journal}
  {Rep. Prog. Phys.}\ }\textbf {\bibinfo {volume} {41}},\ \bibinfo {pages}
  {587} (\bibinfo {year} {1978})}\BibitemShut {NoStop}%
\bibitem [{\citenamefont {Bhattacharyya}(1983)}]{Bhattacharyya1983}%
  \BibitemOpen
  \bibfield  {author} {\bibinfo {author} {\bibfnamefont {K.}~\bibnamefont
  {Bhattacharyya}},\ }\bibfield  {title} {\enquote {\bibinfo {title} {Quantum
  decay and the mandelstam-tamm-energy inequality},}\ }\href@noop {} {\bibfield
   {journal} {\bibinfo  {journal} {J. Phys. A}\ }\textbf {\bibinfo {volume}
  {16}},\ \bibinfo {pages} {2993} (\bibinfo {year} {1983})}\BibitemShut
  {NoStop}%
\bibitem [{\citenamefont {Muga}\ \emph {et~al.}(2009)\citenamefont {Muga},
  \citenamefont {Ruschhaupt},\ and\ \citenamefont {del Campo}}]{MugaBook}%
  \BibitemOpen
  \bibfield  {author} {\bibinfo {author} {\bibfnamefont {J.~G.}\ \bibnamefont
  {Muga}}, \bibinfo {author} {\bibfnamefont {A.}~\bibnamefont {Ruschhaupt}}, \
  and\ \bibinfo {author} {\bibfnamefont {A.}~\bibnamefont {del Campo}},\
  }\href@noop {} {\emph {\bibinfo {title} {Time in Quantum Mechanics, vol.
  2}}}\ (\bibinfo  {publisher} {Springer},\ \bibinfo {address} {London},\
  \bibinfo {year} {2009})\BibitemShut {NoStop}%
\bibitem [{\citenamefont {T\'avora}\ \emph {et~al.}(2016)\citenamefont
  {T\'avora}, \citenamefont {Torres-Herrera},\ and\ \citenamefont
  {Santos}}]{Tavora2016}%
  \BibitemOpen
  \bibfield  {author} {\bibinfo {author} {\bibfnamefont {Marco}\ \bibnamefont
  {T\'avora}}, \bibinfo {author} {\bibfnamefont {E.~J.}\ \bibnamefont
  {Torres-Herrera}}, \ and\ \bibinfo {author} {\bibfnamefont {Lea~F.}\
  \bibnamefont {Santos}},\ }\bibfield  {title} {\enquote {\bibinfo {title}
  {Inevitable power-law behavior of isolated many-body quantum systems and how
  it anticipates thermalization},}\ }\href {\doibase
  10.1103/PhysRevA.94.041603} {\bibfield  {journal} {\bibinfo  {journal} {Phys.
  Rev. A}\ }\textbf {\bibinfo {volume} {94}},\ \bibinfo {pages} {041603}
  (\bibinfo {year} {2016})}\BibitemShut {NoStop}%
\bibitem [{\citenamefont {T\'avora}\ \emph {et~al.}(2017)\citenamefont
  {T\'avora}, \citenamefont {Torres-Herrera},\ and\ \citenamefont
  {Santos}}]{Tavora2017}%
  \BibitemOpen
  \bibfield  {author} {\bibinfo {author} {\bibfnamefont {Marco}\ \bibnamefont
  {T\'avora}}, \bibinfo {author} {\bibfnamefont {E.~J.}\ \bibnamefont
  {Torres-Herrera}}, \ and\ \bibinfo {author} {\bibfnamefont {Lea~F.}\
  \bibnamefont {Santos}},\ }\bibfield  {title} {\enquote {\bibinfo {title}
  {Power-law decay exponents: A dynamical criterion for predicting
  thermalization},}\ }\href {\doibase 10.1103/PhysRevA.95.013604} {\bibfield
  {journal} {\bibinfo  {journal} {Phys. Rev. A}\ }\textbf {\bibinfo {volume}
  {95}},\ \bibinfo {pages} {013604} (\bibinfo {year} {2017})}\BibitemShut
  {NoStop}%
\bibitem [{\citenamefont {Urbanowski}(2009)}]{Urbanowski2009}%
  \BibitemOpen
  \bibfield  {author} {\bibinfo {author} {\bibfnamefont {K.}~\bibnamefont
  {Urbanowski}},\ }\bibfield  {title} {
  %{\selectlanguage {English}
  \enquote
  {\bibinfo {title} {General properties of the evolution of unstable states at
  long times},}\ }\href {\doibase 10.1140/epjd/e2009-00165-x} {\bibfield
  {journal} {\bibinfo  {journal} {Eur. Phys. J. D}\ }\textbf {\bibinfo {volume}
  {54}},\ \bibinfo {pages} {25--29} (\bibinfo {year} {2009})}\BibitemShut
  {NoStop}%
\bibitem [{\citenamefont {Mehta}(1991)}]{MehtaBook}%
  \BibitemOpen
  \bibfield  {author} {\bibinfo {author} {\bibfnamefont {M.~L.}\ \bibnamefont
  {Mehta}},\ }\href@noop {} {\emph {\bibinfo {title} {Random Matrices}}}\
  (\bibinfo  {publisher} {Academic Press},\ \bibinfo {address} {Boston, USA},\
  \bibinfo {year} {1991})\BibitemShut {NoStop}%
\bibitem [{\citenamefont {Schliemann}(2015)}]{Schliemann2015}%
  \BibitemOpen
  \bibfield  {author} {\bibinfo {author} {\bibfnamefont {John}\ \bibnamefont
  {Schliemann}},\ }\bibfield  {title} {\enquote {\bibinfo {title} {Coherent
  quantum dynamics: What fluctuations can tell},}\ }\href {\doibase
  10.1103/PhysRevA.92.022108} {\bibfield  {journal} {\bibinfo  {journal} {Phys.
  Rev. A}\ }\textbf {\bibinfo {volume} {92}},\ \bibinfo {pages} {022108}
  (\bibinfo {year} {2015})}\BibitemShut {NoStop}%
 \bibitem{noteGOE} We note that  for random variables $r_k=x_k^2$,  where $x_k$ come from a Gaussian probability distribution,  Eq.~(\ref{Eq:kappin}) gives $\kappa_\infty^{GOE}= 1/3$, which is the depth obtained for full random matrices from GOE~\cite{Torres2017Philo}.
\bibitem [{\citenamefont {Goussev}\ \emph {et~al.}(2012)\citenamefont
  {Goussev}, \citenamefont {Jalabert}, \citenamefont {Pastawski},\ and\
  \citenamefont {Wisniacki}}]{Goussev2012}%
  \BibitemOpen
  \bibfield  {author} {\bibinfo {author} {\bibfnamefont {A.}~\bibnamefont
  {Goussev}}, \bibinfo {author} {\bibfnamefont {R.~A.}\ \bibnamefont
  {Jalabert}}, \bibinfo {author} {\bibfnamefont {H.~M.}\ \bibnamefont
  {Pastawski}}, \ and\ \bibinfo {author} {\bibfnamefont {D.~A.}\ \bibnamefont
  {Wisniacki}},\ }\bibfield  {title} {\enquote {\bibinfo {title} {Loschmidt
  echo},}\ }\href {\doibase 10.4249/scholarpedia.11687} {\bibfield  {journal}
  {\bibinfo  {journal} {Scholarpedia}\ }\textbf {\bibinfo {volume} {7}},\
  \bibinfo {pages} {11687} (\bibinfo {year} {2012})}\BibitemShut {NoStop}%
\end{thebibliography}
\end{document}